\documentclass[aps,prb,amsmath,amssymb,floatfix,twocolumn]{revtex4-1}
\usepackage{graphicx}
\usepackage{subfigure}
\usepackage{amsmath}
\usepackage{bbold}
\usepackage{slashed}
\usepackage{amssymb}
\usepackage{braket}
\usepackage{array}
\usepackage[colorlinks]{hyperref}
\usepackage{float}
\usepackage[margin=1in]{geometry}
\usepackage{times}

\newcommand{\ncmd}{\newcommand}
\ncmd{\nn}{\nonumber}
\ncmd{\pg}[1]{\textcolor{red}{#1}}
\ncmd{\mbf}[1]{\bs{#1}}
\ncmd{\Lam}{\Lambda}
\ncmd{\lam}{\lambda}
\ncmd{\Gam}{\Gamma}
\ncmd{\gam}{\gamma}
\ncmd{\sig}{\sigma}
\ncmd{\Dl}{\Delta}
\ncmd{\dl}{\delta}
\ncmd{\kap}{\kappa}
\ncmd{\Om}{\Omega}
\ncmd{\om}{\omega}
\ncmd{\mc}{\mathcal}
\ncmd{\eps}{\epsilon}
\ncmd{\veps}{\varepsilon}
\ncmd{\vphi}{\varphi}
\ncmd{\vtheta}{\vartheta}
\ncmd{\note}[1]{{\color{red}{#1}}}
\ncmd{\new}[1]{{\texttt{#1}  } }
\ncmd{\eq}[1]{Eq. \eqref{#1}}
\ncmd{\bs}{\boldsymbol}
\ncmd{\pll}{\parallel}
\ncmd{\dsty}{\displaystyle}

\begin{document}
%\preprint{APS/123-QED}

\title{Spin-charge separation and quantum spin Hall effect of $\beta$-bismuthene}
\author{Alexander C. Tyner$^{1}$ and Pallab Goswami$^{1,2}$}
\affiliation{$^{1}$ Graduate Program in Applied Physics, Northwestern University, Evanston, Illinois, 60208, USA}
\affiliation{$^{2}$ Department of Physics and Astronomy, Northwestern University, Evanston, Illinois, 60208, USA}

\date{\today}

\begin{abstract}
Field theory arguments suggest the possibility of $\mathbb{Z}$-classification of quantum spin Hall effect with magnetic flux tubes, that cause separation of spin and charge degrees of freedom, and pumping of spin or Kramers pair. However, the \emph{proof of principle} demonstration of spin-charge separation is yet to be accomplished for realistic, \emph{ab initio} band structures of spin-orbit-coupled materials, lacking spin-conservation law. In this work, we perform thought experiments with magnetic flux tubes on $\beta$-bismuthene to demonstrate spin-charge separation, and quantized pumping of spin for three insulating states that can be accessed by tuning filling fractions. With a combined analysis of momentum-space topology and real-space response, we identify important role of topologically non-trivial bands, supporting even integer winding numbers, which cannot be inferred from symmetry-based indicators. Our work sets a new standard for prediction of two-dimensional, quantum spin-Hall materials, based on precise bulk invariant and universal topological response.
\end{abstract}

\maketitle

\section{Introduction} Chern insulators~\cite{Laughlin1981,TKNN1982,PhysRevLett.61.2015} and quantum spin Hall (QSH) insulators~\cite{Kane2005,bernevig2006quantum,MurakamiQSHBi,konig2007quantum,FuKane,Roy2009} are two prominent examples of two-dimensional (2D) topological phases of matter. For crystalline materials, they respectively arise due to the existence of net Chern number ($\mathfrak{C}_{GS} \in \mathbb{Z}$)~\cite{TKNN1982,PhysRevLett.61.2015,Niu1985} and net relative or spin Chern number ($\mathfrak{C}_{R,GS} \in \mathbb{Z}$)~\cite{Kane2005,bernevig2006quantum} for completely occupied bands. The Chern number $\mathfrak{C}_{GS}$ of time-reversal-symmetry ($\mathcal{T}$) breaking materials describes quantized Abelian Berry flux ($2 \pi \mathfrak{C}_{GS}$) through 2D Brillouin zone (BZ) and it can be calculated using TKNNY formula~\cite{TKNN1982}, and by imposing twisted boundary condition (TBC) in real space~\cite{Niu1985}. Furthermore, the topological response (pumping of electric charge $\Delta Q = e \mathfrak{C}_{GS}$) can be directly probed by inserting a magnetic flux tube carrying flux $\phi$, and adiabatically tuning $\phi$ between $0$ and $\phi_0=h/e$~\cite{Laughlin1981}. 

In contrast to $\mathfrak{C}_{GS}$, $\mathfrak{C}_{R,GS}$ describes quantized, non-Abelian Berry flux $2 \pi \mathfrak{C}_{R,GS}$ through 2D BZ. The definition of non-Abelian Berry flux for spin-orbit-coupled materials has many subtleties due to the absence of continuous spin rotation symmetry or spin conservation law. Kane and Mele formulated $\mathbb{Z}_2$-classification of QSH effect of $\mathcal{T}$-symmetric systems, i.e., odd vs. even integer distinction of $\mathfrak{C}_{R,GS}$~\cite{Kane2005}. For generic $\mathcal{T}$-symmetric systems, the $\mathbb{Z}_2$ invariant $(-1)^{\mathfrak{C}_{R,GS}}$ can be calculated from the gauge-invariant spectrum of Wilson loops (Wannier charge centers) of non-Abelian Berry connection~\cite{yu2011equivalent,Soluyanov2011,alexandradinata2014wilson,Z2pack}. Furthermore, for inversion-symmetric materials, it can be easily identified from a symmetry-based indicator, which is the product of parity eigenvalues at time-reversal-invariant momentum points~\cite{FuKane}. 

To go beyond $\mathbb{Z}_2$-classification of $\mathfrak{C}_{R,GS}$ without relying on spin and momentum conservation laws, various authors have considered the role of generalized TBCs~\cite{ShengHaldane,QiTBC,qi2008topological}. However, the insertion of spin-gauge flux and the implementation of spin-dependent TBC requires detailed understanding of underlying basis states and the mechanism of violation of spin-conservation law. Hence, its application has practical limitation. 

To overcome this challenge, Qi and Zhang~\cite{SpinChargeSCZ}, and Ran \emph{et. al.}~\cite{VishwanathPiFlux} proposed diagnosis of $\mathbb{Z}_2$ QSH states with magnetic flux tubes (i.e., gauging of conserved quantity). Employing 4-band models of QSH states with $|\mathfrak{C}_{R,GS}|= 1$, they have shown that a flux tube, carrying half of flux quantum $\phi = \frac{\phi_0}{2}$ ($\mathcal{T}$-invariant, $\pi$ flux) binds two degenerate, zero-energy, mid-gap states. At half-filling, one of these modes is occupied, and the ground state exhibits $2$-fold-degeneracy ($SU(2)$-doublet). Consequently, the flux tube remains charge-neutral, and carries spin quantum number $\pm \frac{1}{2}$. When both modes are occupied (empty), the flux tube carries electric charge $-e$ ($+e$), and spin quantum number $0$ ($SU(2)$-singlets). Such states can be accessed by doping insulators with one electron (hole). This solitonic mechanism of \emph{spin-charge separation} (SCS) is similar to what is known for polyacetelene~\cite{PhysRevD.13.3398,ssh1979} and topologically ordered, correlated systems~\cite{kivelson1987}. When $\mathcal{T}$ is broken by generic values of $\phi$, the bound modes and the half-filled ground state become non-degenerate. But the flux tube continues to bind spin and no electric charge. By adiabatically tuning $\phi$ from $0$ to $\phi_0$, quantized pumping of spin (one Kramers-pair) can be observed [see Appendix~\ref{AppA}]. 

%%%%%%%%%%%%%%%%%%%%%%%%%%%%
\begin{figure*}[t]
\centering
\subfigure[]{
\includegraphics[scale=0.1]{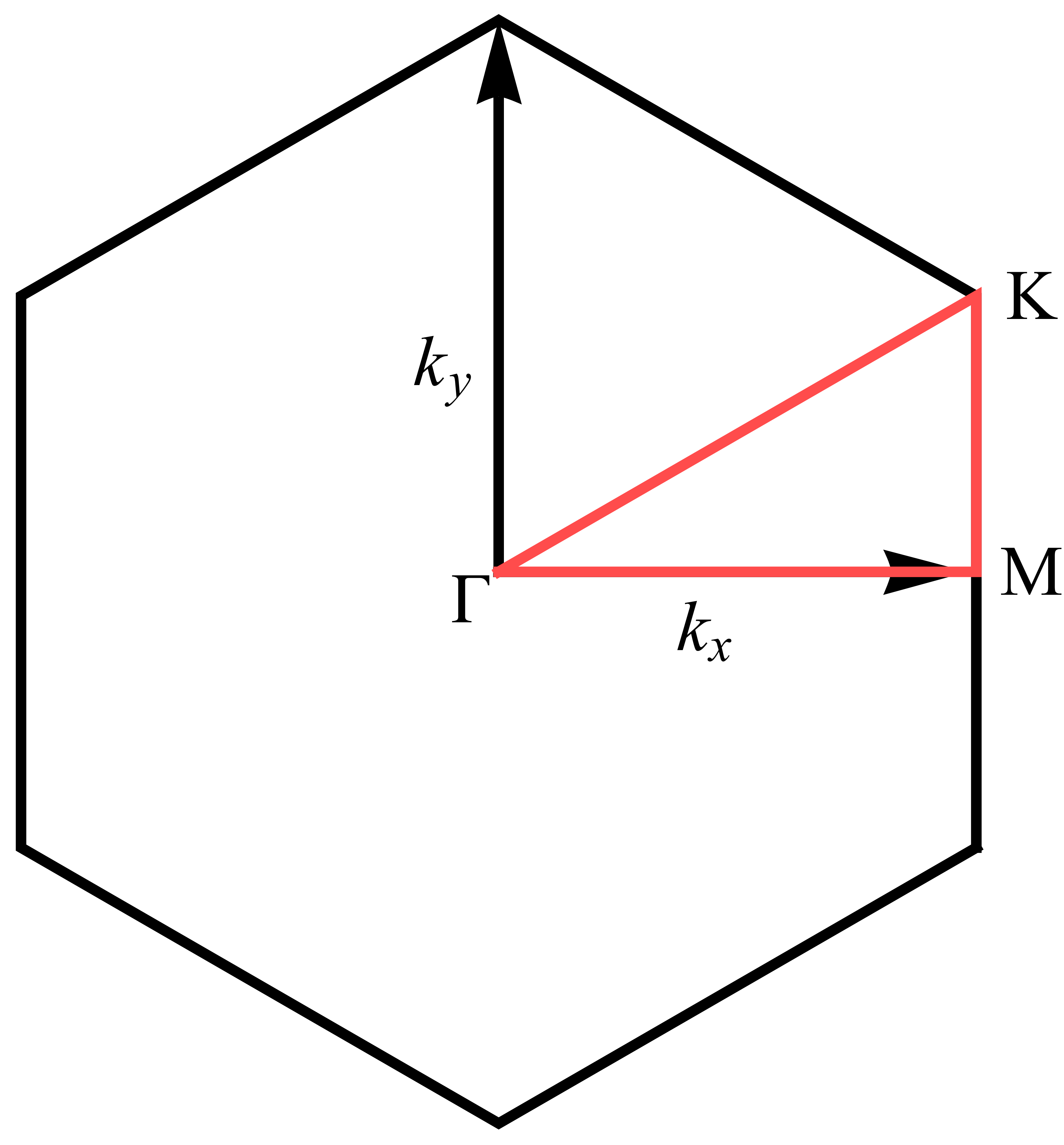}
\label{fig:BiBZ}}
\subfigure[]{
\includegraphics[scale=0.65]{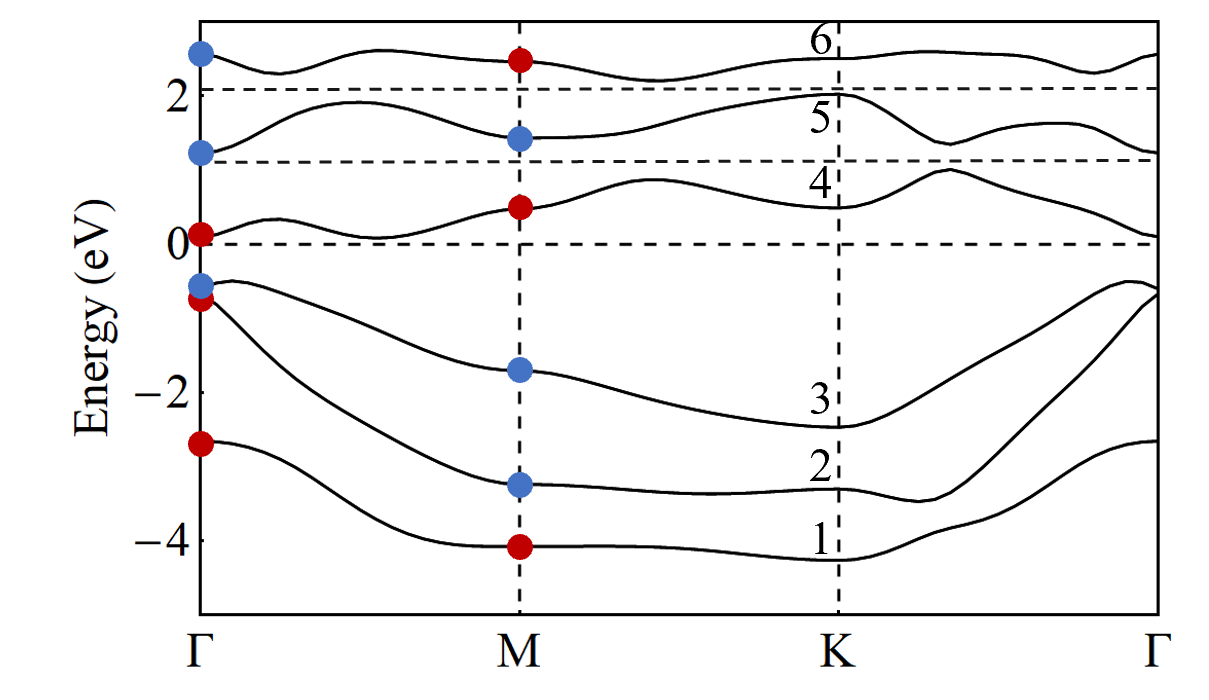}
\label{fig:BiML}}
\subfigure[]{
\includegraphics[scale=0.85]{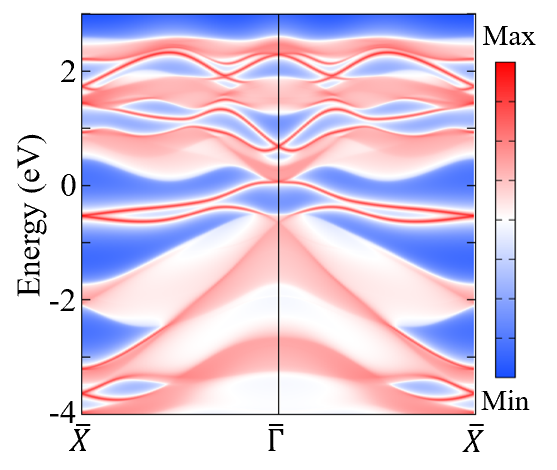}
\label{fig:BiSSY}}
\subfigure[]{
\includegraphics[scale=0.85]{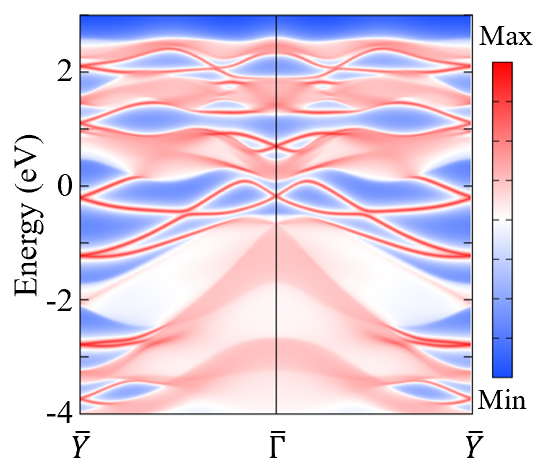}
\label{fig:BiSSX}}
\caption{(a) Hexagonal Brillouin zone of $\beta$-bismuthene. (b) Band structure along high-symmetry path $\Gamma-M-K-\Gamma$ and energies are measured with respect to a reference value $E_0=0$. Bands are numbered according to their energies at $\Gamma$ point, such that $E_{n}(\Gamma)<E_{n+1}(\Gamma)$. Parity eigenvalues $\pm 1$ at time-reversal-invariant-momentum points are denoted by red and blue dots, respectively. The dashed lines correspond to three representative values of Fermi energy, tuned in direct band gaps, leading to $1/2$-, $2/3$-, and $5/6$- filled insulators. They support non-trivial $\mathbb{Z}_2$-invariant $\nu_{0,GS}=1$. (c,d) Spectral density on the surface under open boundary conditions along (c) y-axis and (d) x-axis. The mid-gap edge-modes, connecting bulk valence and conduction bands imply first-order topology of insulating states. }
\end{figure*}
%%%%%%%%%%%%%%%%%%%%%%%%%%%%

%%%%%%%%%%%%%%%%%%%%%%%%%%%%
\begin{figure}
    \centering
    \includegraphics[width=8.5cm]{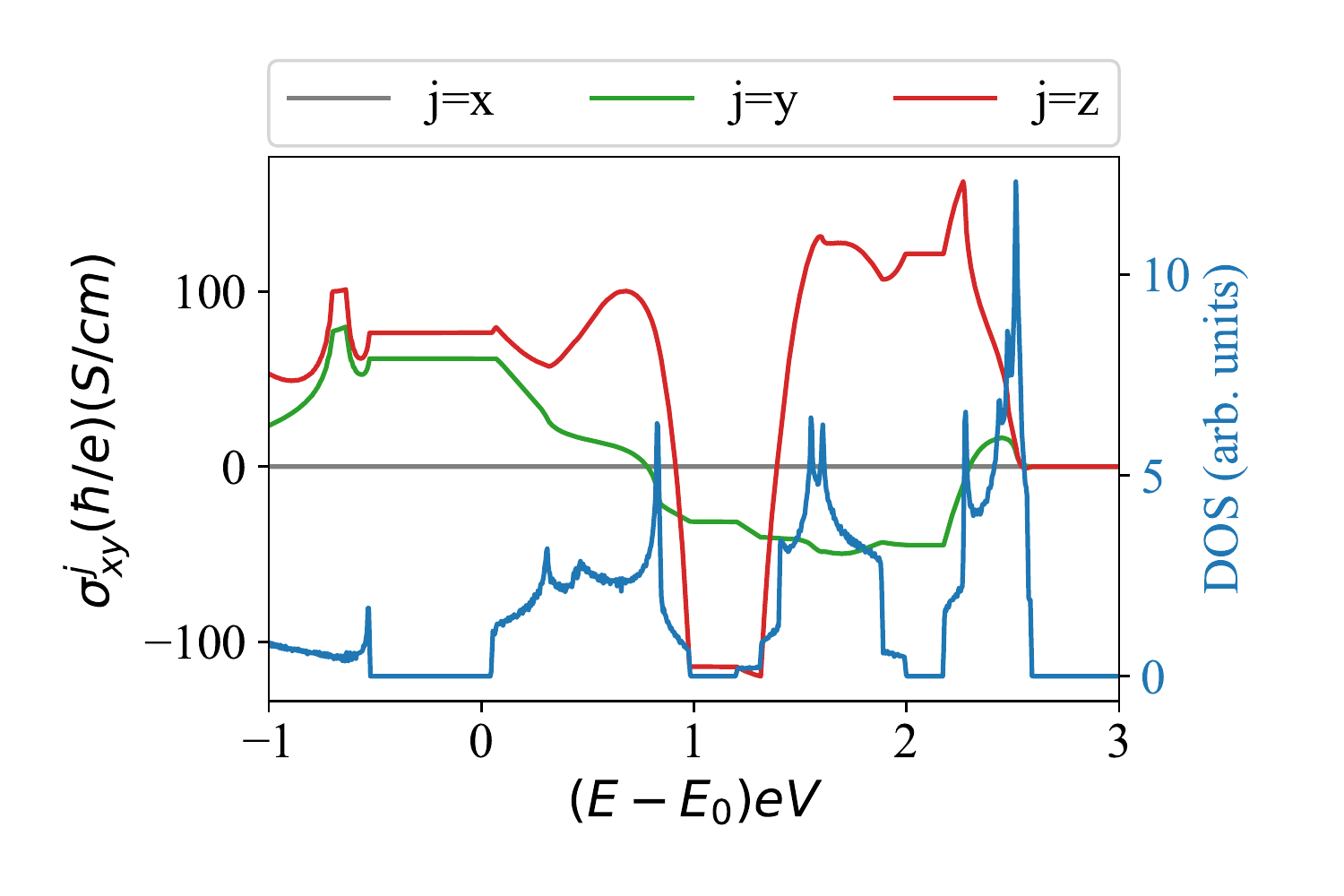}
    \caption{Spin Hall conductivity $\sigma^{j}_{xy}$ of $\beta$-bismuthene from first principles calculations, as a function of energy. Three insulating states can be identified from direct gaps in density of states. In addition to showing plateau-like features, $\sigma^{z}_{xy}$ shows sharp change of sign, when bands 4 and 5 become occupied. While these results cannot capture precise topological properties, they indicate topological non-triviality of these $\mathbb{Z}_2$-trivial bands.}
    \label{fig:BiSpinHall}
\end{figure}
%%%%%%%%%%%%%%%%%%%%%%%%%%%%

While Refs.~\onlinecite{SpinChargeSCZ,VishwanathPiFlux} and subsequent works~\cite{Wang_2010,slager2012,MESAROS2013977} advanced conceptual understanding of QSH effect, they relied on idealized models of decoupled Chern insulators, carrying opposite Chern  numbers. Due to the underlying $U(1)$ spin-rotation symmetry, these models admit $\mathbb{Z}$-classification of $\mathfrak{C}_{R,GS}=\mathfrak{C}_{GS,\uparrow}=-\mathfrak{C}_{GS,\downarrow}$, and the spin-Hall conductivity $\sigma^z_{xy} = 2 \mathfrak{C}_{R,GS}$. The SCS for such systems would be governed by $SU(2 |\mathfrak{C}_{R,GS}|)$-multiplets [see Appendix~\ref{AppA}]. Since $\mathfrak{C}_{R,GS}$ for decoupled models is easily calculated, the analysis of SCS only serves academic interest. 

For realistic band structures of spin-orbit-coupled materials, various crystalline-symmetry allowed hybridization terms destroy $U(1)$ spin-conservation law. Since there is no simple theoretical framework for computing $\mathfrak{C}_{R,GS}$ of such systems, the demonstration of SCS would allow unambiguous diagnosis of $|\mathfrak{C}_{R,GS}|$ or $\mathbb{N}$-classification of QSH insulators. Moreover, $\mathbb{Z}$-classification can be accomplished by measuring spin expectation values, during the process of spin-pumping.
Recently, we have addressed the stability of SCS for topologically non-trivial planes of 3D Dirac semimetals ($4$-band model)~\cite{tyner2020topology}, and $3$-fold symmetric planes ($8$-band model)~\cite{tynerWitteneffect} of octupolar topological insulators~\cite{Benalcazar61}. These models support SCS respectively controlled by $SU(2)$ and $SU(4)$ multiplets. Moreover, the quantized pumping of spin occurs even in the absence spin-rotation symmetry and gapless, helical edge-states. 

Encouraged by these results, in this work, we perform \emph{proof of principle} demonstration of SCS for realistic, \emph{ab initio} band structures. For concreteness, we focus on a single $(111)$-bilayer of elemental bismuth (Bi), also known as $\beta$-bismuthene, as a suitable material platform. The analysis of SCS will be guided by the calculation of gauge-invariant magnitudes of relative Chern numbers of constituent bands ($|\mathfrak{C}_{R,n}|$ for $n$-th band)~\cite{tyner2021quantized}. Thus, the importance of $\mathbb{Z}_2$-trivial bands, possessing even integer winding number ($\mathfrak{C}_{R,j}=2s_j \neq 0$) will be critically addressed. We also present a brief contrasting study of $\beta$-antimonene, which does not exhibit SCS.

\section{Band topology of bismuthene} 
Symmetry-based topological classification of 3D Bi has dramatically evolved over past fifteen years. The band structure of Bi was initially classified as topologically trivial, with strong $\mathbb{Z}_2$ TI index $\nu_{0,GS}=0$~\cite{FuKane,teo2008surface}. Now it is identified as a higher-order, topological crystalline insulator with strong $\mathbb{Z}_4$ index $\kappa_{1,GS}= 2$~\cite{schindler2018higher,hsu2019topology,aggarwal2021evidence}. In contrast to this, the ground state of $\beta$-bismuthene is known to be a $\mathbb{Z}_2$ QSH insulator~\cite{WadaBi,MurakamiQSHBi,drozdov2014one,BiFilmIto,reis2017bismuthene,bieniek2017stability,TakayamBi}, supporting helical edge states~\cite{WadaBi,MurakamiQSHBi,drozdov2014one}.

The crystal structure of $\beta$-bismuthene is described by buckled honeycomb layers with space group $P6/mcc$. The material supports $\mathcal{T}$ and space-inversion ($\mathcal{P}$) symmetries, leading to the two-fold Kramers degeneracy of all energy bands throughout the hexagonal BZ of Fig.~\ref{fig:BiBZ}. Using the crystal structure and lattice constants from Ref.~\onlinecite{mounet2018two}, the \emph{ab initio} band structure has been calculated with Quantum Espresso~\cite{QE-2009,QE-2017,QE-2020}. 
The band structure along high-symmetry path $\Gamma-M-K-\Gamma$ is shown in Fig.~\ref{fig:BiML}. As these bands are well separated from other bands, an accurate $12$-band, Wannier tight-binding (TB) model has been constructed from $p_{x,y,z}$ orbitals from each layer. We have included spin-orbit coupling for all calculations and utilized a 40 x 40 x 1 Monkhorst-Pack grid of $\bs{k}$-points and a plane wave cutoff of 100 Ry. The model construction and topological analysis are performed with Wannier90 and Z2pack~\cite{Pizzi2020,Z2pack,Soluyanov2011}.

When the Fermi level is tuned inside direct gaps between bands (i) 3 and 4, (ii) 4 and 5, and (iii) 5 and 6, we find three insulators at filling fractions $1/2$, $2/3$, and $5/6$. From the parity eigenvalues shown in Fig.~\ref{fig:BiML}, we find that only bands $2$ and $6$ possess non-trivial $\mathbb{Z}_2$-index $\nu_{0,n}=1$, and all three insulators admit non-trivial $\mathbb{Z}_2$-index $\nu_{0,GS}=1$. 
The results of edge-states calculations, using iterative Greens function method~\cite{sancho1985highly} and Wannier Tools~\cite{WU2017} are displayed in Figs.~\ref{fig:BiSSX} and ~\ref{fig:BiSSY}. All three insulators support gapless edge modes, which can cross the Fermi level $2$ or $6$ times~\cite{WadaBi}. Whether $|\mathfrak{C}_{R,GS}|=1$ or $3$ cannot be determined from edge-spectrum. 

It is instructive to compute spin Hall conductivity, following the current state-of-the-art of computational materials science,~\cite{QSHWannier,destraz2020magnetism,tsirkin2021high} and the results are shown in Fig.~\ref{fig:BiSpinHall}. Due to the non-conservation of spin, this method cannot capture quantization of spin Hall effect. But it provides rough guidance for understanding qualitative properties of three insulators. Notice that $\sigma^z_{xy}$ displays plateau-like features, when $E$ is tuned in direct band gaps, and changes sign when $\mathbb{Z}_2$-trivial bands 4 and 5 become occupied. Are bands $4$ and $5$ topologically trivial? \begin{table}
\def\arraystretch{1.5}
	\begin{tabular}{|c|c|c|c|}
		\hline
		\begin{tabular}{c}Band \\
		index $n$\end{tabular} & \begin{tabular}{c} $3$-fold eigen-  \\ values $C_{3,n}$ \end{tabular} & \begin{tabular}{c} $\mathbb{Z}_2$-index\\ $\nu_{0,n}$ \end{tabular} & \begin{tabular}{c} Relative Chern  \\ number $|\mathfrak{C}_{R,n}| $ \end{tabular} \\
		\hline
		$1$ & $e^{\pm i \frac{\pi}{3}}$ & $0$ & 0 \\
		\hline
		$2$ & $e^{\pm i \frac{\pi}{3}}$ & $1$ & 1\\
		\hline
		$3$ & $e^{\pm i \frac{\pi}{3}}$ & $0$ & 0 \\
		\hline
		$4$ & $e^{\pm i \pi}$ & 0 & 2 \\
		\hline
		$5$ & $e^{\pm i \frac{\pi}{3}}$ & 0 & 2 \\
		\hline
		$6$ & $e^{\pm i \pi}$ & 1 & 1 \\
		\hline
	\end{tabular}
  \caption{Natural number classification of relative Chern numbers of constituent Kramers-degenerate bands of $\beta$-bismuthene. The $3$-fold rotation eigenvalues and $\mathbb{Z}_2$-indices are listed for convenience. The symmetry data of rotation and parity eigenvalues are insufficient to distinguish between bands possessing $|\mathfrak{C}_{R,n}|=0$ and $2$.}
    \label{tab:Indicators}
    \end{table}

This question can be conclusively answered by identifying the magnitude of relative Chern number of Kramers-degenerate bands ($|\mathfrak{C}_{R,n}|$) by computing in-plane Wilson loops of $SU(2)$ Berry connection~\cite{tyner2021quantized}. The results are displayed in Table~\ref{tab:Indicators} and the details of calculations are presented in Appendix~\ref{AppB}. We see that bands $4$ and $5$ carry even integer invariants. While they do not change odd integer classification of $\mathfrak{C}_{R, GS}$, they can change the magnitude and the sign of $\mathfrak{C}_{R, GS}$. Table~\ref{tab:Indicators} suggests the following possibilities: (i) $|\mathfrak{C}_{R, GS}| = 1$, (ii) $|\mathfrak{C}_{R, GS}|=1,3$, (iii) $|\mathfrak{C}_{R, GS}|=1,3,5$, respectively for $1/2$, $2/3$, and $5/6$ filled insulators. While it is possible to compute signed $\mathfrak{C}_{R,n}$ by adding a small time-reversal symmetry breaking field, we will resolve uncertainties with SCS. 

%%%%%%%%%%%%%%%%%%%%%%%%%%%% Through a generalization of non-Abelian Stokes theorem, eigenvalues of in-plane Wilson loop are related to the gauge-invariant magnitude of $SU(2)$ Berry flux, enclosed by the loop. Adhering to underlying rotational symmetry, we have calculated in-plane Wilson loops, using $C_3$-symmetric contours and  The details of calculation are presented in Appendix. 
%%%%%%%%%%%%%%%%%%%%%%%%%%%%

\begin{figure*}[t]
\centering
\subfigure[]{
\includegraphics[scale=0.49]{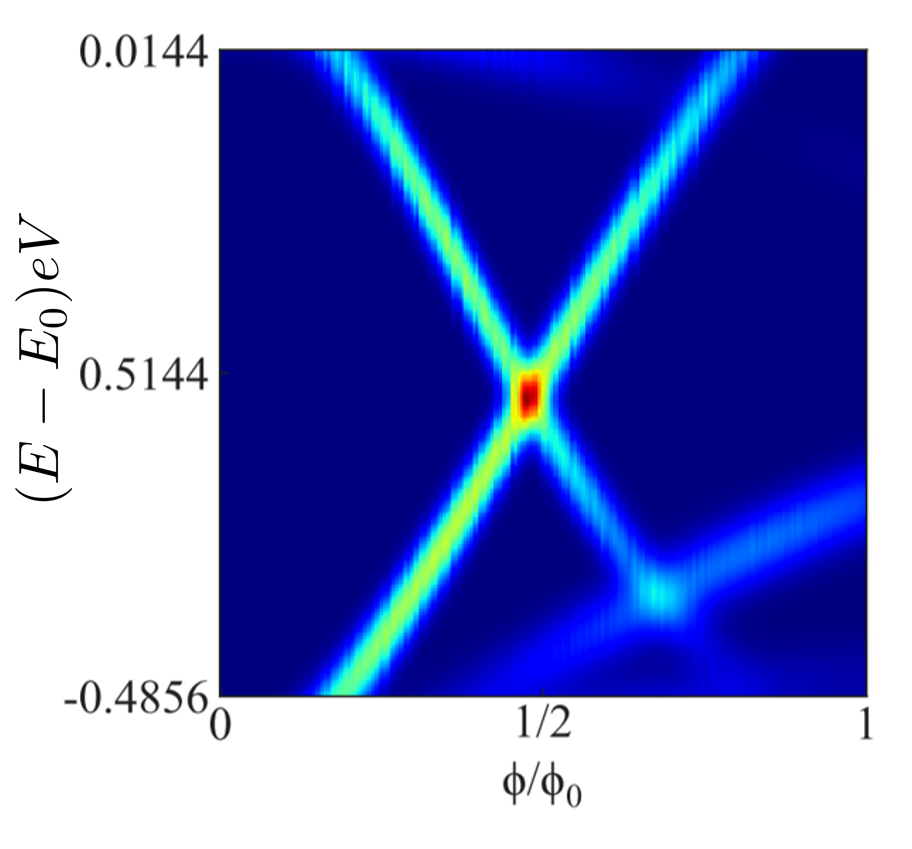}
\label{fig:BiFluxHalf}}
\subfigure[]{
\includegraphics[scale=0.5]{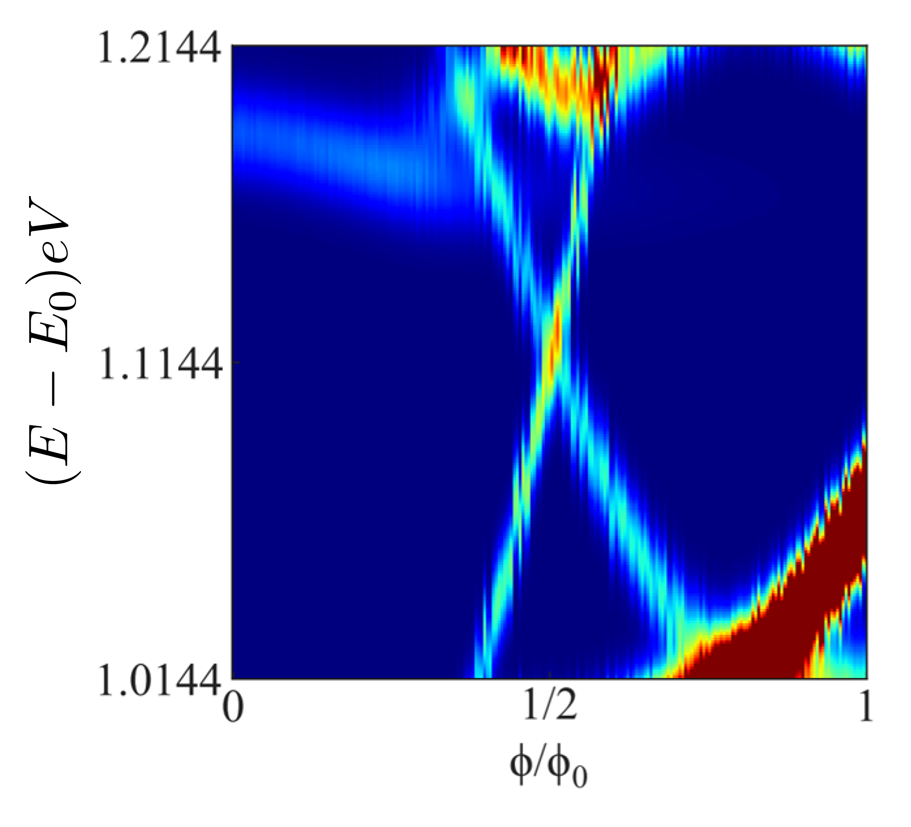}
\label{fig:BiFluxTwoThird}}
\subfigure[]{
\includegraphics[scale=0.51]{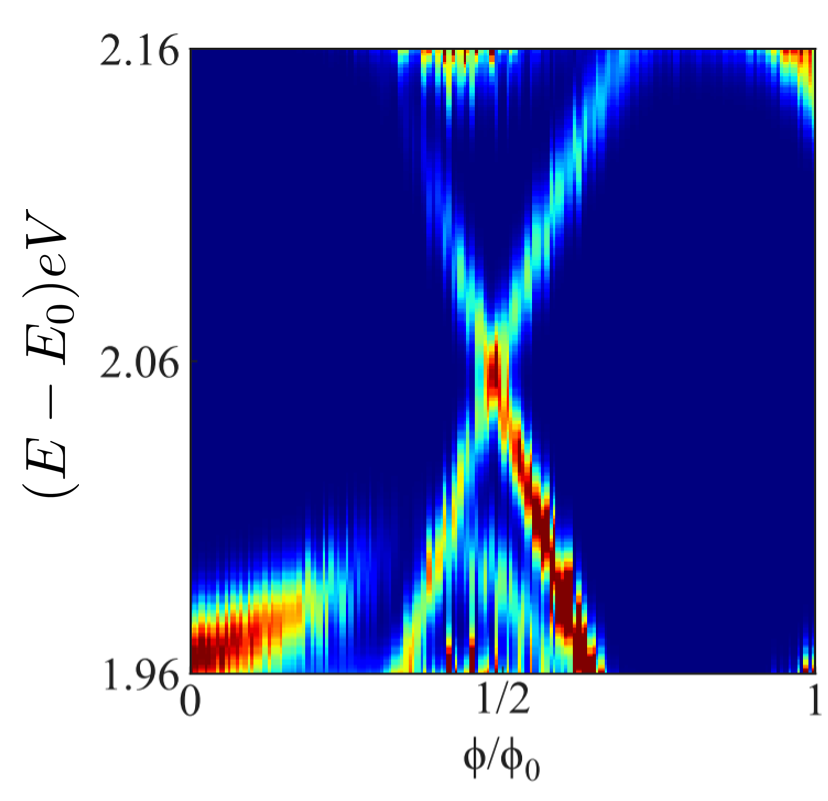}
\label{fig:Flux56}}
\subfigure[]{
\includegraphics[scale=0.345]{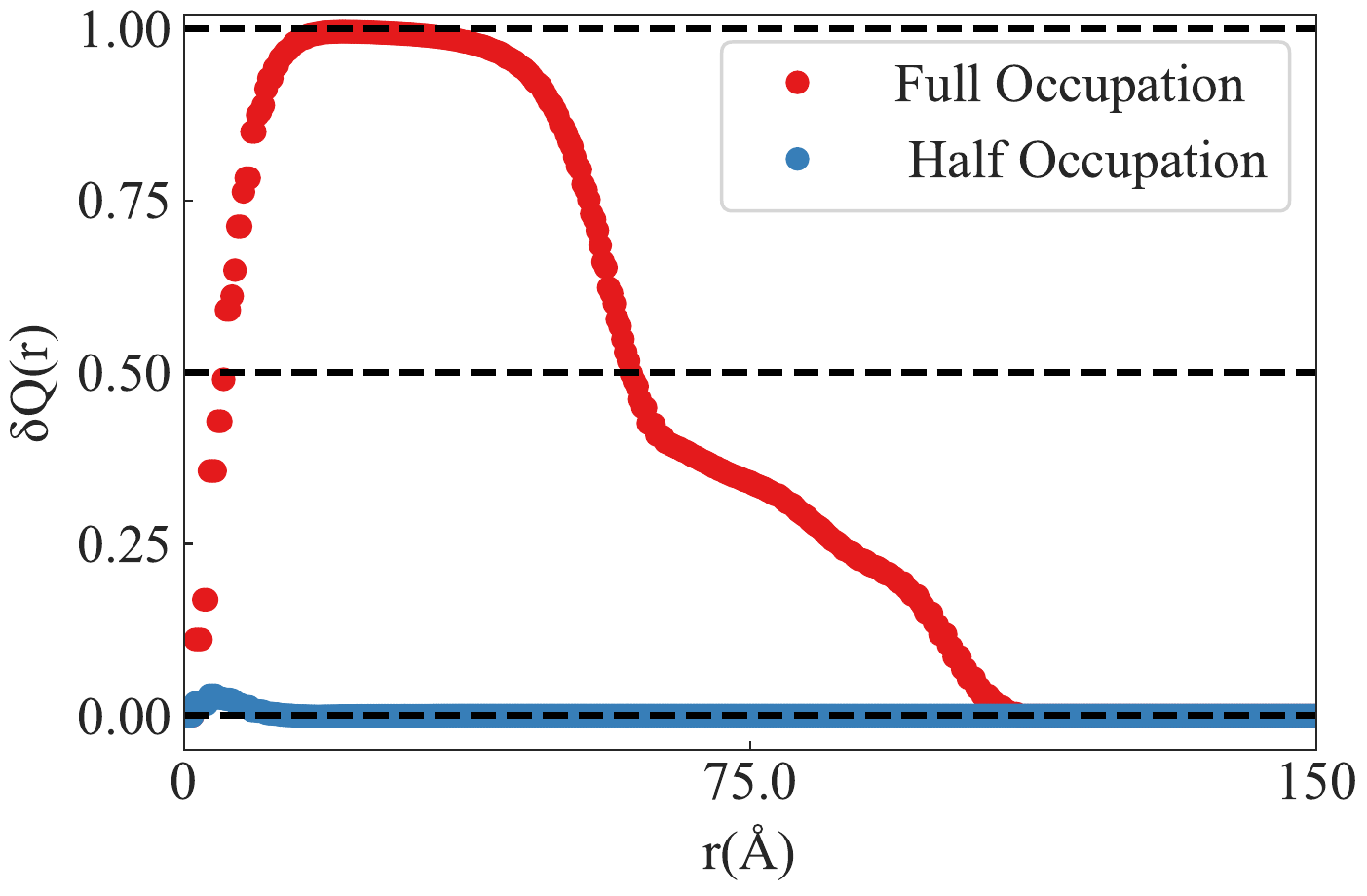}
\label{fig:BiICHalf}}
\subfigure[]{
\includegraphics[scale=0.345]{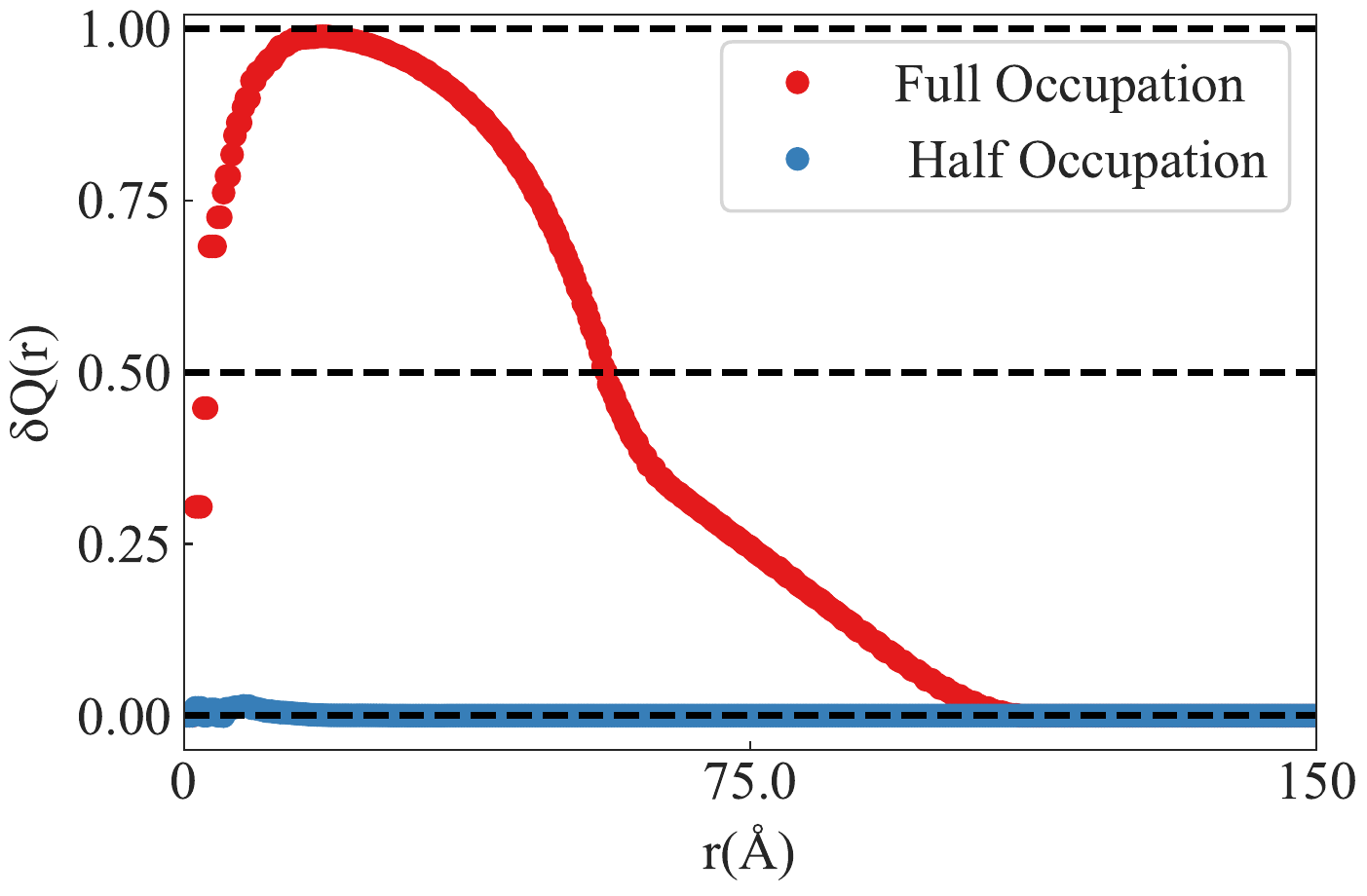}
\label{fig:BiICTwoThird}}
\subfigure[]{
\includegraphics[scale=0.345]{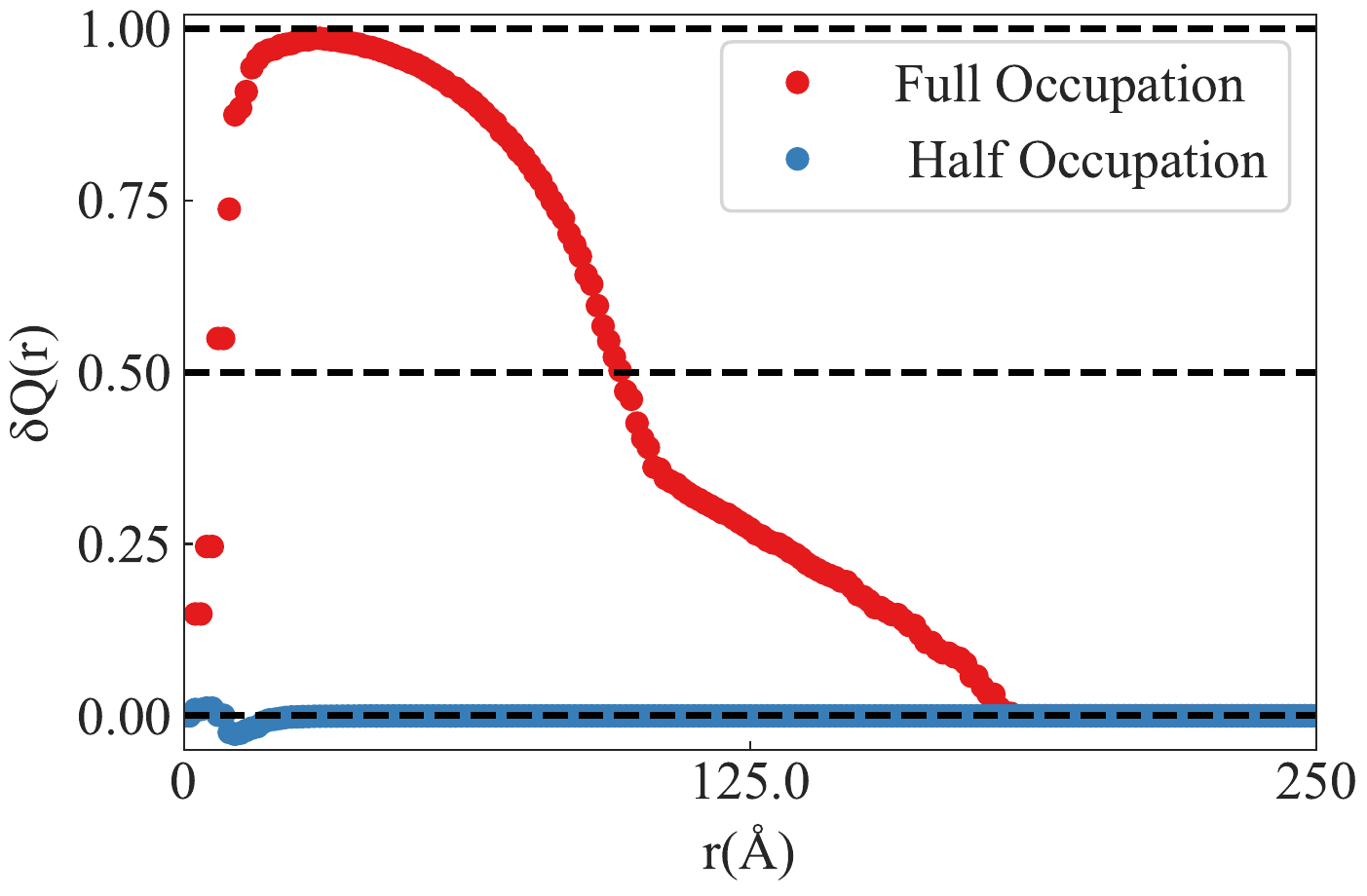}
\label{fig:IC56}}
\subfigure[]{
\includegraphics[scale=0.355]{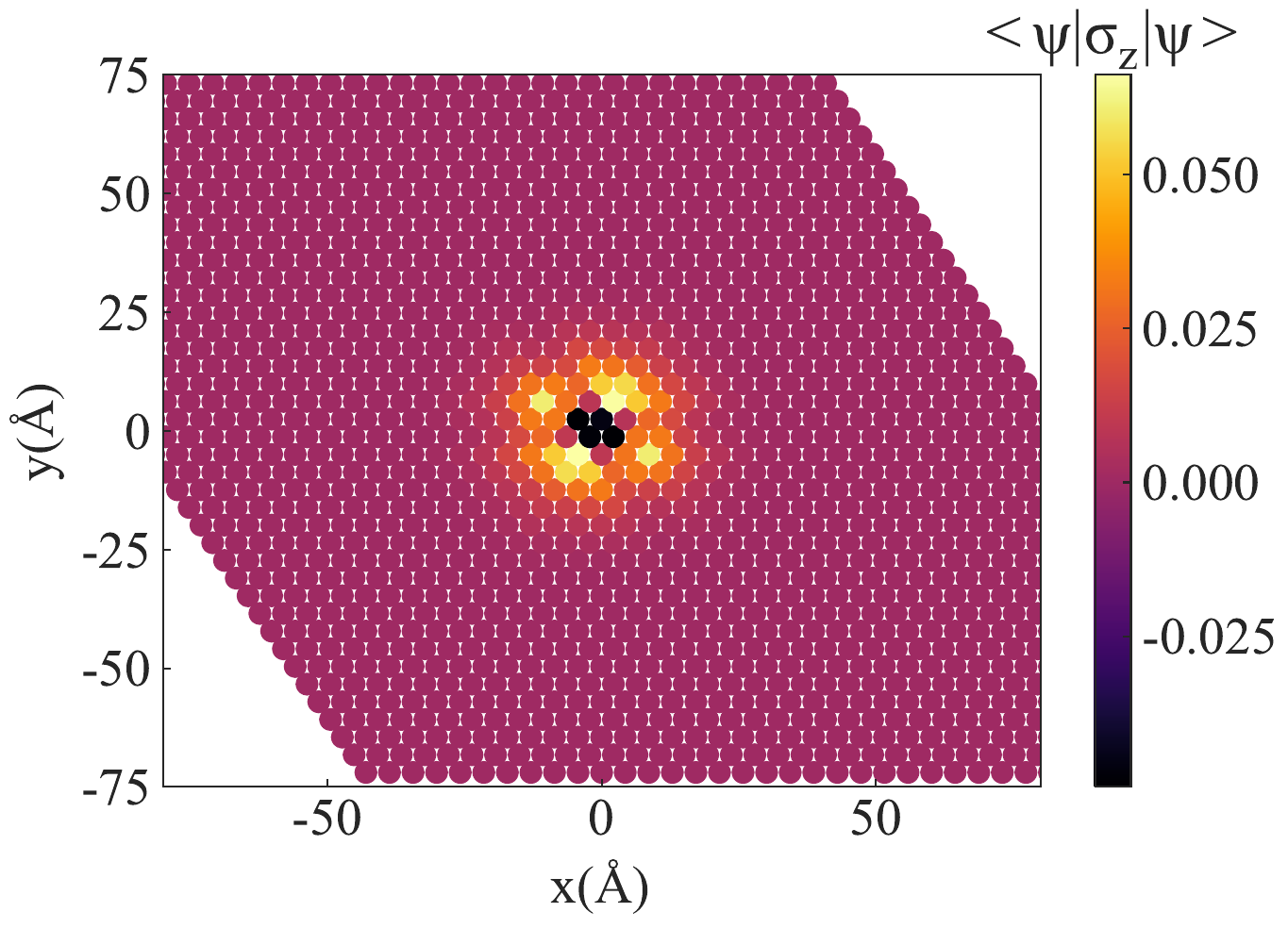}
\label{fig:BiSz12}}
\subfigure[]{
\includegraphics[scale=0.355]{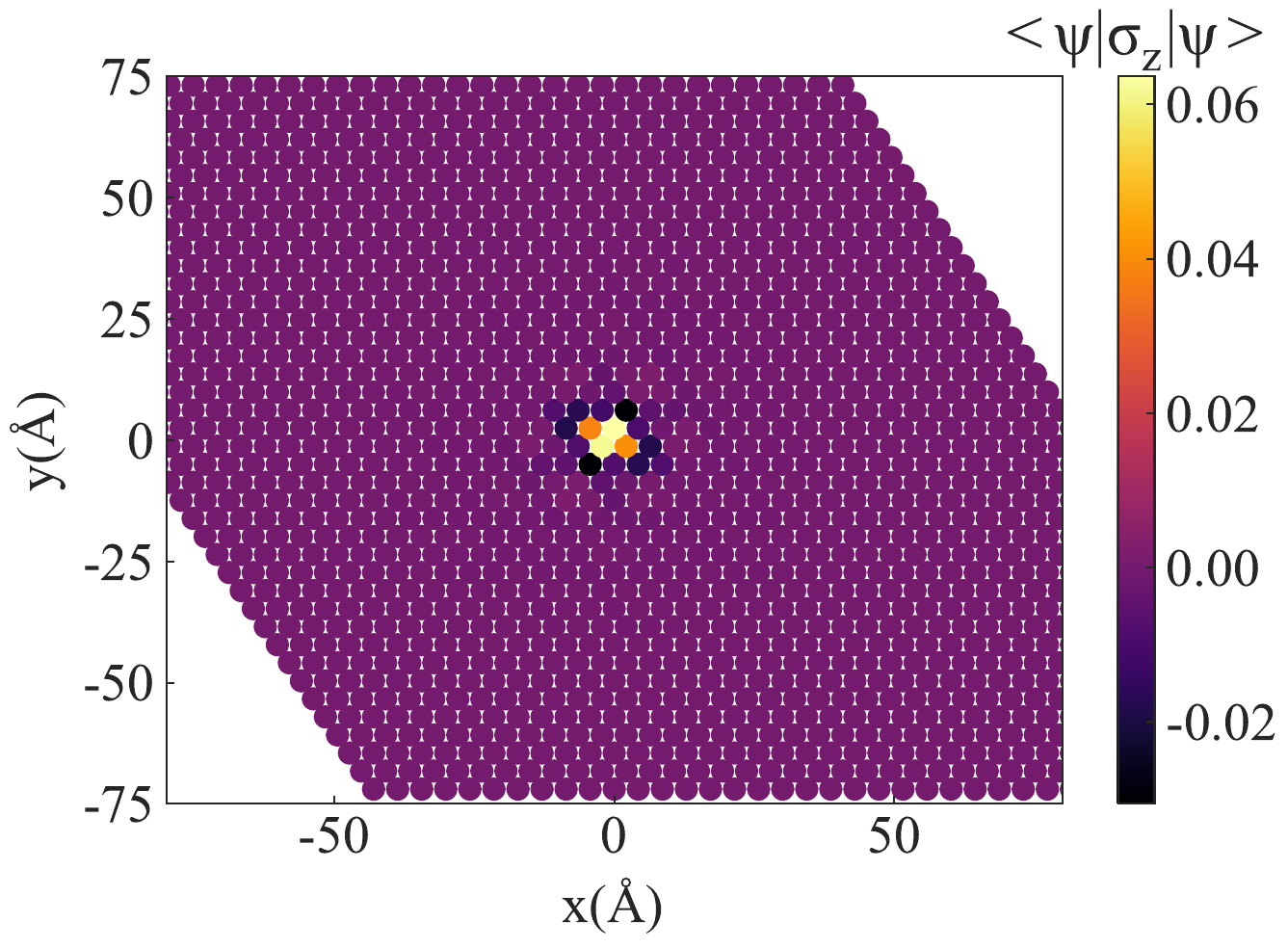}
\label{fig:BiSz23}}
\subfigure[]{
\includegraphics[scale=0.355]{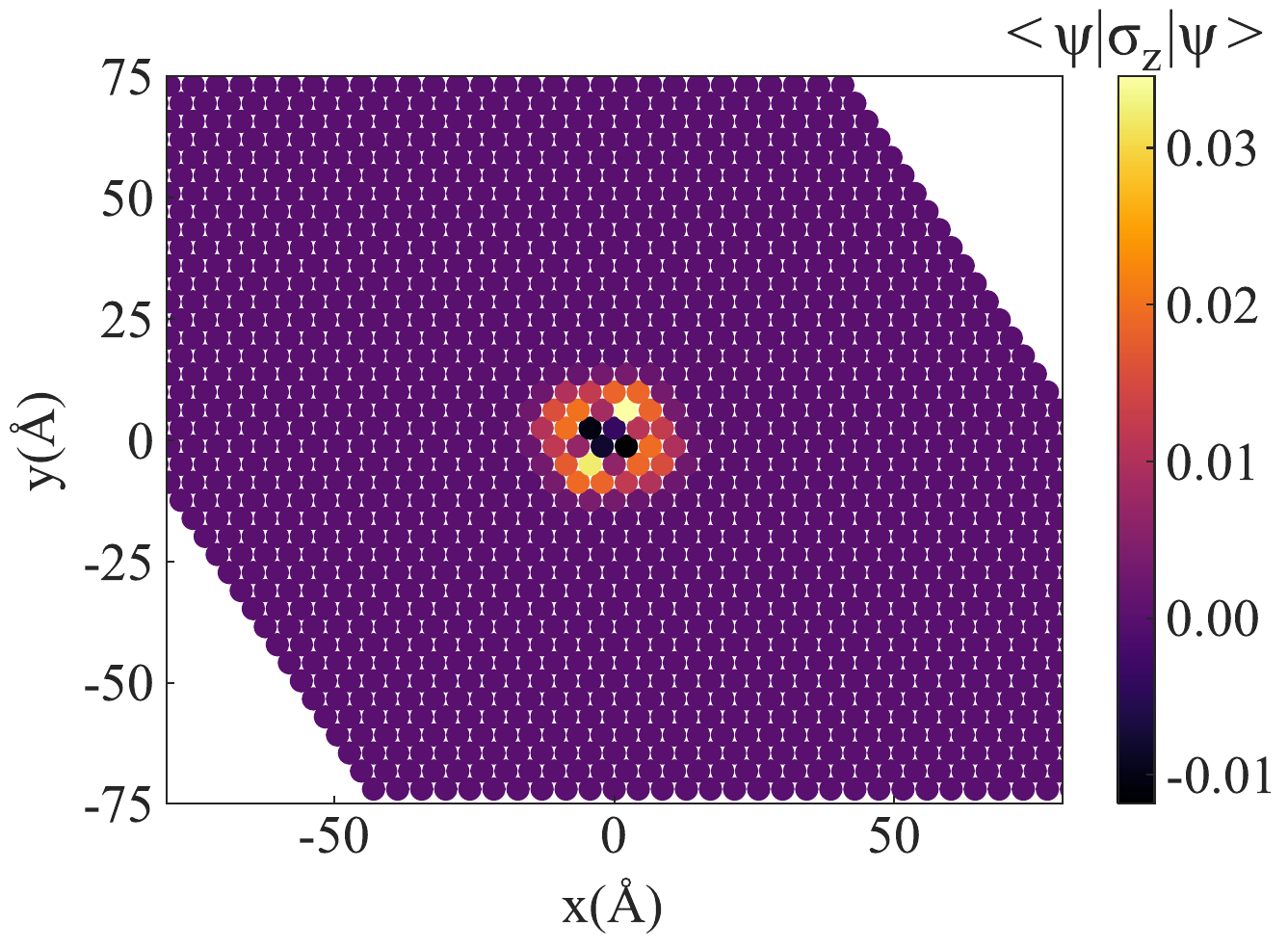}
\label{fig:BiSz56}}
\caption{Spin-charge separation in $\beta$-bismuthene. (a)-(c) Local density of states on flux tube, as a function of $\phi/\phi_0$, with energies scanned in the vicinity of band gaps. For time-reversal-invariant flux $\phi=\phi_0/2$, two-fold degenerate bound states occur precisely at $1/2$, $2/3$, and $5/6$ filling fractions. For any generic value of flux, time-reversal-symmetry is broken and bound states become non-degenerate. As $\phi$ is tuned from $0$ to $\phi_0$, one Kramers-pair is pumped, revealing $|\mathfrak{C}_{R,GS}|=1$ for all three insulators. Induced electric charge $\delta Q(r)$ (in units of $-e$) on flux tube, as a function of radial distance $r$ from the flux tube, for (d) $1/2$-filled, (e) $2/3$-filled, (f) $5/6$-filled insulators, with $\phi/\phi_0 =1/2$. When the bound states are half-filled (fully occupied), the maximum value of induced charge saturates to quantized value $0$ ($-e$). The results remain unchanged for $\phi/\phi_0 \neq 1/2$. The spin expectation values of unoccupied bound modes for $\phi=(\frac{1}{2}-\epsilon)\phi_{0}$, and $\epsilon=10^{-3}$ for (g) $1/2$-filled, (h) $2/3$-filled, (i) $5/6$-filled insulators. The profile of spin polarization for $2/3$-filled insulator is opposite to those for $1/2$- and $5/6$- filled insulators. }
\end{figure*}

%%%%%%%%%%%%%%%%%%%%%%%%%%%
\section{Spin charge separation}
To study real-space topological response we insert a flux tube at the center of 2D system. The hopping matrix element $H_{ij}$ connecting lattice sites $\bs{r}_i$ and $\bs{r}_j$ is modified to $H_{ij} \; e^{i \phi_{ij}}$. Working with Coulomb gauge, we define the Peierls phase factor 
\begin{equation}
\phi_{ij}= \frac{\phi}{\phi_0} \int_{\bs{r}_i}^{\bs{r}_j} \frac{\hat{z} \times \bs{r}}{\bs{r}^2} \cdot d\bs{l}.
\end{equation}
We first perform exact diagonalization of gauged Hamiltonian for $24 \times 24$ unit cells, under periodic boundary conditions (PBC), yielding $N=6912$ eigenstates. When the total number of electrons $N_e=\frac{N}{2}, \frac{2N}{3}, \frac{5N}{6}$, and $\phi=\phi_{0}/2$, we find two-fold-degenerate mid-gap states bound to the flux tube, leading to two-fold degeneracy of ground states [see Figs.~\ref{fig:BiFluxHalf}-\ref{fig:BiFluxTwoThird}]. When $\phi \neq \phi_0/2$, flux tube breaks time-reversal-symmetry and the degeneracy of bound states (ground state) is lifted. By holding $N_e$ fixed at commensurate values, and varying $\phi$ from $0$ to $\phi_0$, we observe pumping of one Kramers pair.~\cite{SpinChargeSCZ} 
\par 
Next we compute the induced electric charge on flux-tube for $\phi=\phi_{0}/2$. This calculation is done in two steps.~\cite{rosenberg2010witten} For a given number of electrons, by summing over \emph{all occupied states}, we evaluate the area charge densities $\sigma_1(\bs{r}_i, N_e)$, and $\sigma_0(\bs{r}_i, N_e)$, respectively in the presence and absence of flux tube. The induced charge density is defined as $\delta \sigma (\bs{r}_i, N_e) = \sigma_1(\bs{r}_i, N_e)-\sigma_0(\bs{r}_i, N_e)$, and the total induced charge within a circle of radius $r$, centered at the flux tube is determined from $\delta Q(r,N_e)=\sum_{|\bs{r}_i|<r}\delta \sigma(\bs{r}_i, N_e)$. In order to achieve sufficient numerical accuracy, induced charge calculations are performed for a system size of $60 \times 60$ unit cells, yielding $N=43,200$ eigenstates. The results are displayed in Figs.~\ref{fig:BiICHalf}-\ref{fig:IC56}.
When $N_e=\frac{N}{2}, \frac{2N}{3}, \frac{5N}{6}$, one of the degenerate mid-gap modes is occupied (half-occupation of mid-gap states), and we find $\delta Q(r,N_e) =0$, which remains unchanged for generic values of $\phi$. For $N_e= \frac{N}{2} \pm 1, \frac{2N}{3} \pm 1, \frac{5N}{6} \pm 1$, the bound modes become completely occupied ($+$) and empty ($-$), and the maximum values of $\delta Q (r, N_e)$ saturate to quantized results $\mp e$, respectively. %Notice that $\delta Q( r \to 0) \to 0$, as normalizable wave functions vanish at $r=0$.

Therefore, we can conclude that each non-trivial insulator supports $|\mathfrak{C}_{R, GS}|=1$, which can only be consistent with the following assignments of signed relative Chern numbers 
\begin{eqnarray}
&&(\mathfrak{C}_{R, 1},\mathfrak{C}_{R, 2},\mathfrak{C}_{R, 3},\mathfrak{C}_{R, 4},\mathfrak{C}_{R, 5},\mathfrak{C}_{R, 6}) \nn \\
 &&= \pm (0, 1,0,-2,+2,-1),
\end{eqnarray}
defined with respect to a global spin quantization axis for all bands. When bands $4$ and $5$ are occupied, $\mathfrak{C}_{R,GS}$ will change sign. This assertion can be further substantiated by evaluating expectation values of spin operators $\mathbb{1}_{6 \times 6} \otimes \boldsymbol \sigma \equiv \boldsymbol \sigma$ for bound states. We have computed expectation values for $\phi=(\frac{1}{2} - \epsilon)\phi_{0}$ and $\epsilon \rightarrow 0^{+}$, such that the bound states are infinitesimally split in energy. The expectation values for $\sigma_{x,y}$ are negligibly small. As occupied and unoccupied modes support opposite signs for $\bra{\psi_{n}} \sigma_{z} \ket{\psi_{n}}$, we are only showing the results for unoccupied modes in Figs.~\ref{fig:BiSz12}-\ref{fig:BiSz56}. All signs become reversed for $\phi=(\frac{1}{2} +\epsilon)\phi_{0}$, as a consequence of spin-pumping. Therefore, flux tube for $1/2$-, $2/3$-, $5/6$- filled insulators respectively support $+$, $-$, $+$ signs for $\bra{\psi_{n}} \sigma_{z} \ket{\psi_{n}}$, when $\phi<\phi_0 /2$. %We further elaborate on this relationship for analytically known models in Appendix. % in . While the observed deviation in the pattern of the spin expectation value is to be expected, prediction of a change of sign at each lattice site is qualitatively observed. This is quantitatively substantiated calculating $\hat{S}_{z}=\sum_{n}  \bra{\psi_{n}} \sigma_{z} \ket{\psi_{n}}$, where the sum is over all lattice sites; at $1/2$ and $2/3$ filling $\hat{S}_{z}$ is positive and negative respectively. As a result of this analysis we are able to compile the data presented in Tab. \eqref{tab:Flux_SpinHall}, detailing the non-Abelian flux supported by for band, $n$, as well as the corresponding quantized spin Hall conductivity when bands $1-n$ are occupied. The magnitude and relative sign of the data is conclusive, however we are unable to rule out a global sign change to all invariants.

\section{Conclusions}
In summary, we have shown that bilayer bismuth is a suitable platform for studying spin charge separation as a universal topological response of quantum spin Hall insulators. The combined analysis of non-Abelian Berry phase in momentum space and real space topological response clearly identify non-trivial topology of bands that carry even integer winding numbers. Topology of such bands are not easily detected by symmetry-based indicators. In Appendix~\ref{AppC} we also analyze bulk topology of $\beta$-antimonene, which supports $\mathbb{Z}_2$-trivial ground state. With first principles based calculations of spin Hall conductivity and insertion of magnetic flux tube we show that $\beta$-antimonene is not a quantum spin Hall insulator ( i.e., $\mathfrak{C}_{R,GS}=0$).

In this work, we have only gauged a conserved quantity (electric charge) to identify the presence or absence of spin-pumping. This can be reliably used for many candidate materials for quantum spin Hall effect. Guided by the results of this work, one can further pursue insertion of spin-gauge flux (gauging of non-conserved quantity) to demonstrate pumping of electric charge ($\delta Q= 2 e \mathfrak{C}_{R,GS}$), which directly tracks signed relative/spin Chern number. Due to technical subtleties and numerical cost of such calculations, such thought experiments on real materials would be reported in a future work.

\appendix

\begin{figure}
    \centering
  \includegraphics[scale=0.33]{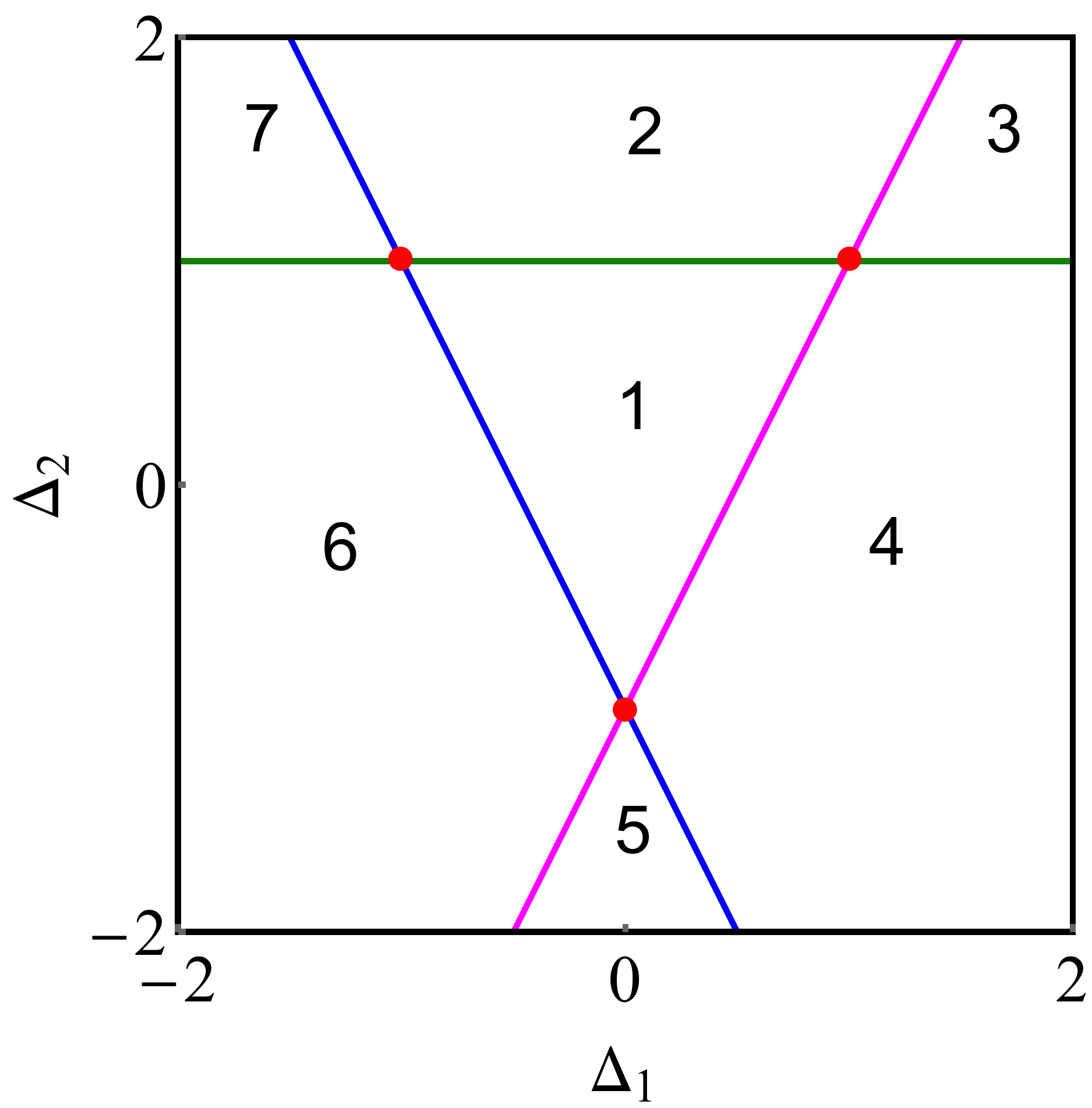}
   \caption{Phase diagram of four-band model for $M=1$. Parity eigenvalues and bulk winding numbers are listed in Table~\ref{tab2}. Along blue, magenta, and green lines, the bulk band gap can close at $\Gamma$, $M$, and $X$ points, respectively. Red dots denote multi-critical points.}
    \label{fig:phasediagram}
\end{figure}

\begin{table}
\def\arraystretch{1.5}
	\begin{tabular}{|c|c|c|}
		\hline
		Phase & \begin{tabular}{c}
			 Parity eigenvalues \\ $(\delta_\Gamma, \delta_M, \delta_X)$
		\end{tabular} & $\mathfrak{C}_{R,GS}$\\
		\hline 
		$1$ & $(-1, -1,-1)$ & $0$\\
		\hline
		$2$ & $(-1, -1,+1)$ & $2 \; \text{sgn}(t)$\\
		\hline
		$3$ & $(-1, +1,+1)$ & $ \text{sgn}(t)$\\
		\hline
		$4$ & $(-1, +1,-1)$ & $- \text{sgn}(t)$\\
		\hline
		$5$ & $(+1, +1,-1)$ & $-2 \; \text{sgn}(t)$\\
		\hline
		$6$ & $(+1, -1,-1)$ &$- \text{sgn}(t)$\\
		\hline
		$7$ & $(+1, -1,+1)$ & $ \text{sgn}(t)$\\
		\hline
		\end{tabular}
\caption{Patterns of parity eigenvalues and bulk winding numbers for various phases of Fig.~\ref{fig:phasediagram}.} \label{tab2}
\end{table}

%%%%%%%%%%%%%%%%%%%%%%%%%%%%
\begin{figure*}
    \centering
    \includegraphics[scale=0.4]{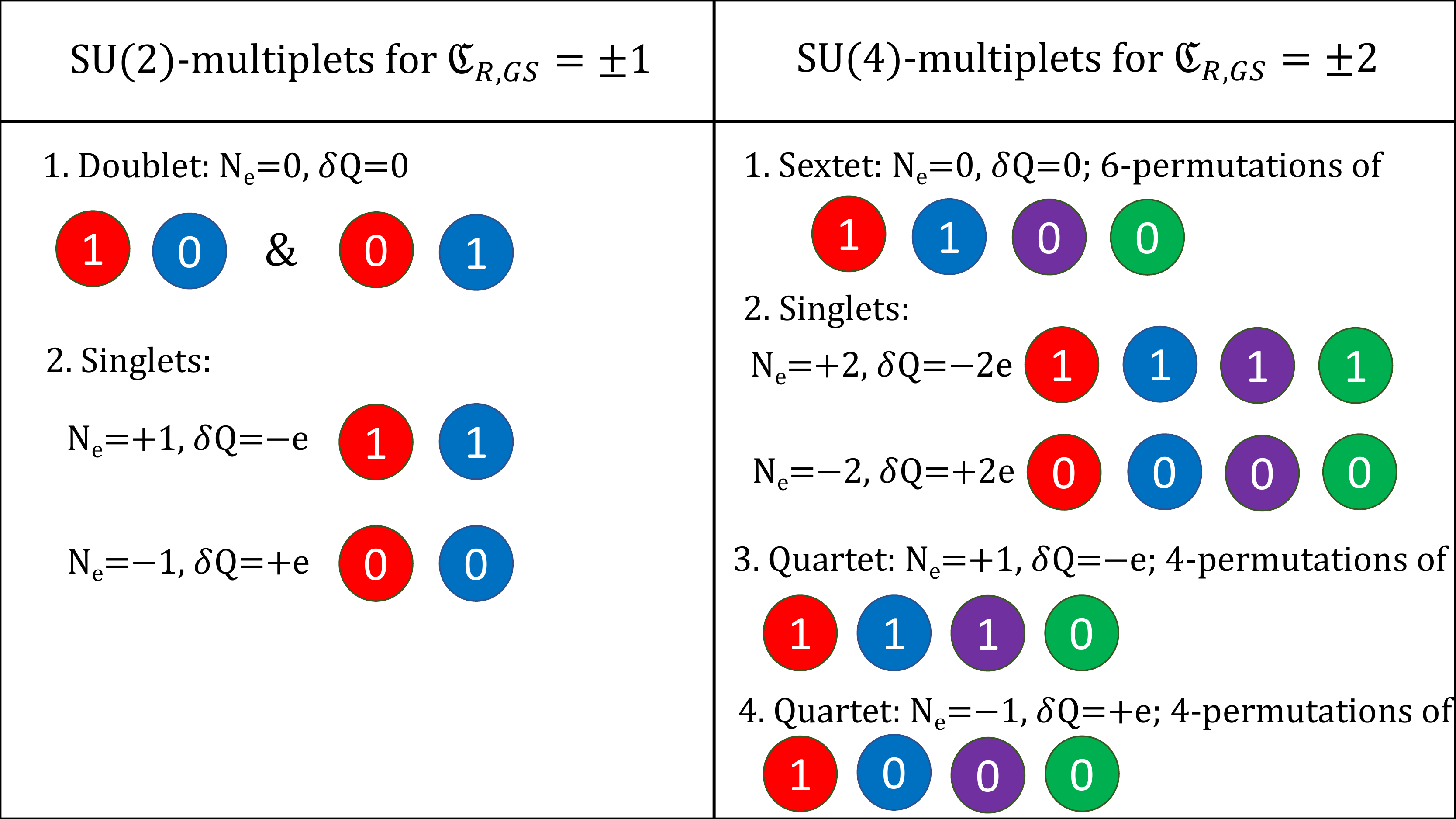}
    \caption{Schematic of spin-charge separation and induced quantum numbers for magnetic $\pi$ flux tube. The occupation number of mid-gap states is denoted by $0$ and $1$. For half-filled systems, the number of added electrons $N_e=0$, and the induced electric charge $\delta Q=0$. By adding one electron or hole one can access $\delta Q= \mp e$ on flux tube for $\mathfrak{C}_{R,GS}=\pm 1$. Additional charge quantum numbers are found for $\mathfrak{C}_{R,GS}=\pm 2$.}
    \label{fig:presentation}
\end{figure*}
%%%%%%%%%%%%%%%%%%%%%%%%%%%%

\begin{figure*}[t]
\centering
\subfigure[]{
\includegraphics[scale=0.4]{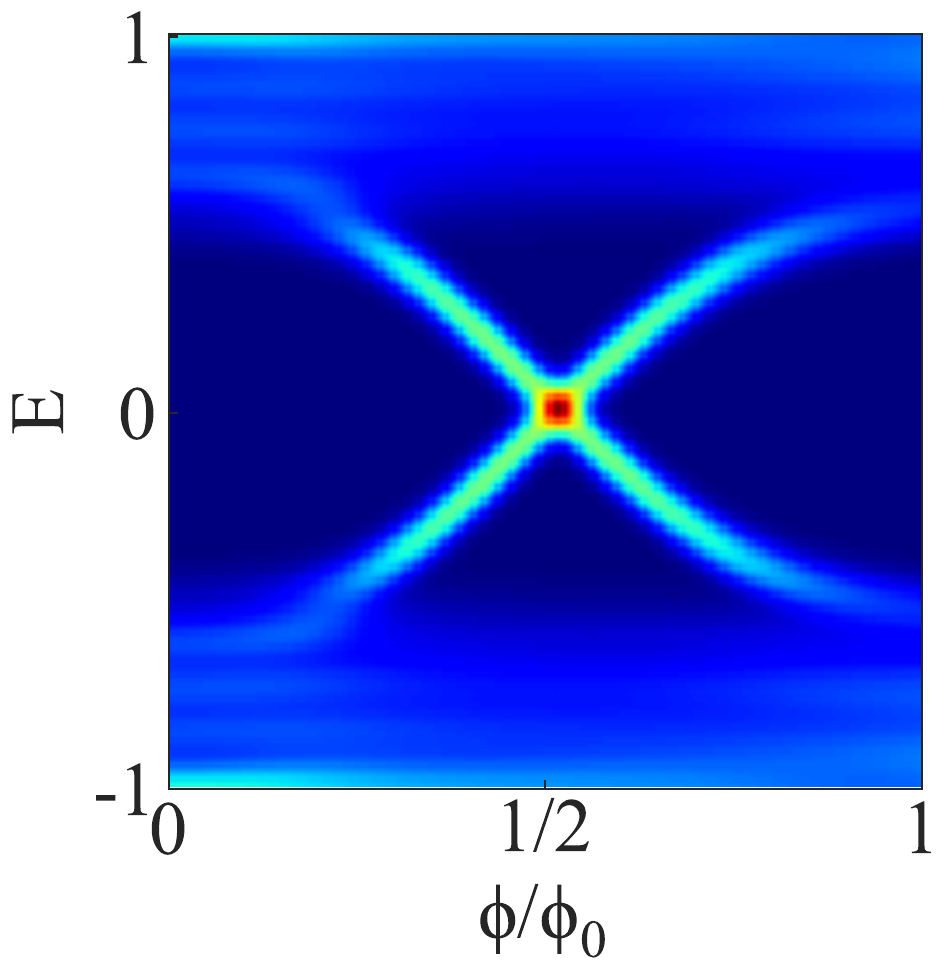}
\label{fig:BHZCharge}}
\subfigure[]{
\includegraphics[scale=0.41]{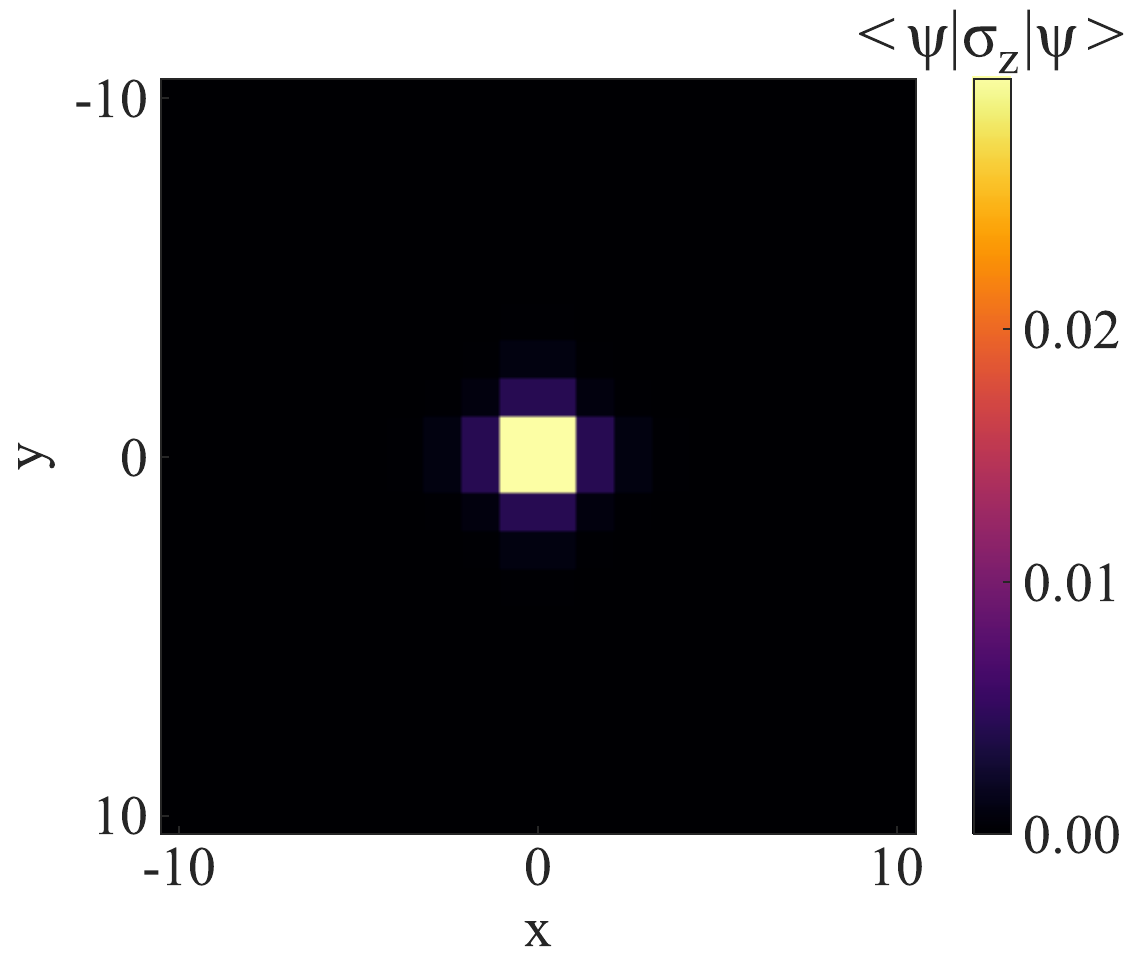}
\label{fig:BHZpos}}
\subfigure[]{
\includegraphics[scale=0.41]{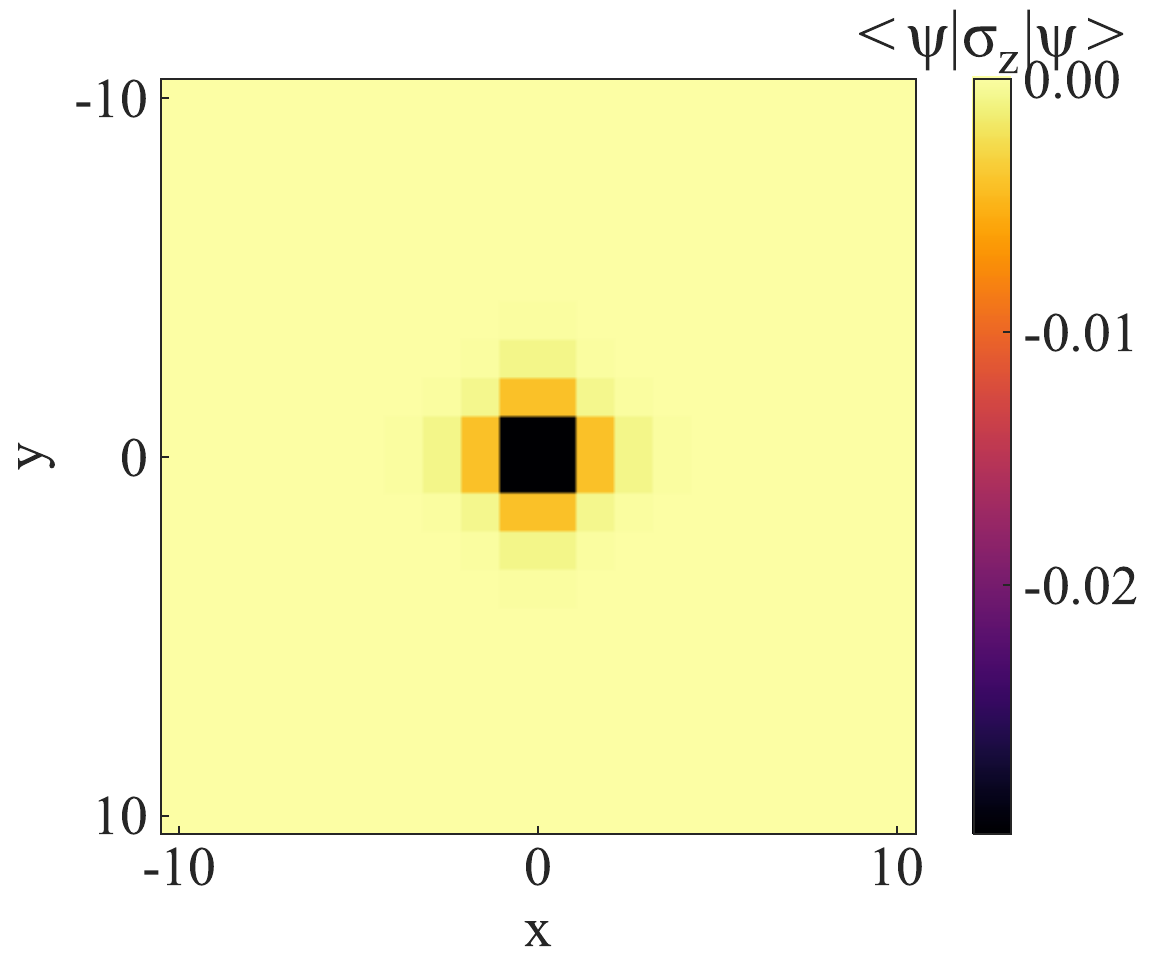} 
\label{fig:BHZneg}}
\caption{Spin-charge separation for Phase 3 and Phase 4, with $t>0$. (a) Local density of states on flux tube for both phases shows pumping of one Kramers pair. Spin expectation value $\bra{\psi_{n}} \sigma_{z} \ket{\psi_{n}}(x,y)$ of unoccupied bound mode, when $\phi=(\frac{1}{2}-\epsilon)\phi_{0}$, and $\epsilon =10^{-2}$, for (b) Phase 3, (c) Phase 4. The spin density on flux tube tracks $\text{sgn}(\mathfrak{C}_{R,GS})$. %Identical spin-pumping (d) and the sign reversal of spin expectation value are found for (e) $M=-3/2$, and (f) $M=+3/2$, for the model of Eq.~\ref{eq:BHZpert}, which lacks spin-rotation symmetry.
}
\label{4band}
\end{figure*}

\begin{figure*}[t]
\centering
\subfigure[]{
\includegraphics[scale=0.4]{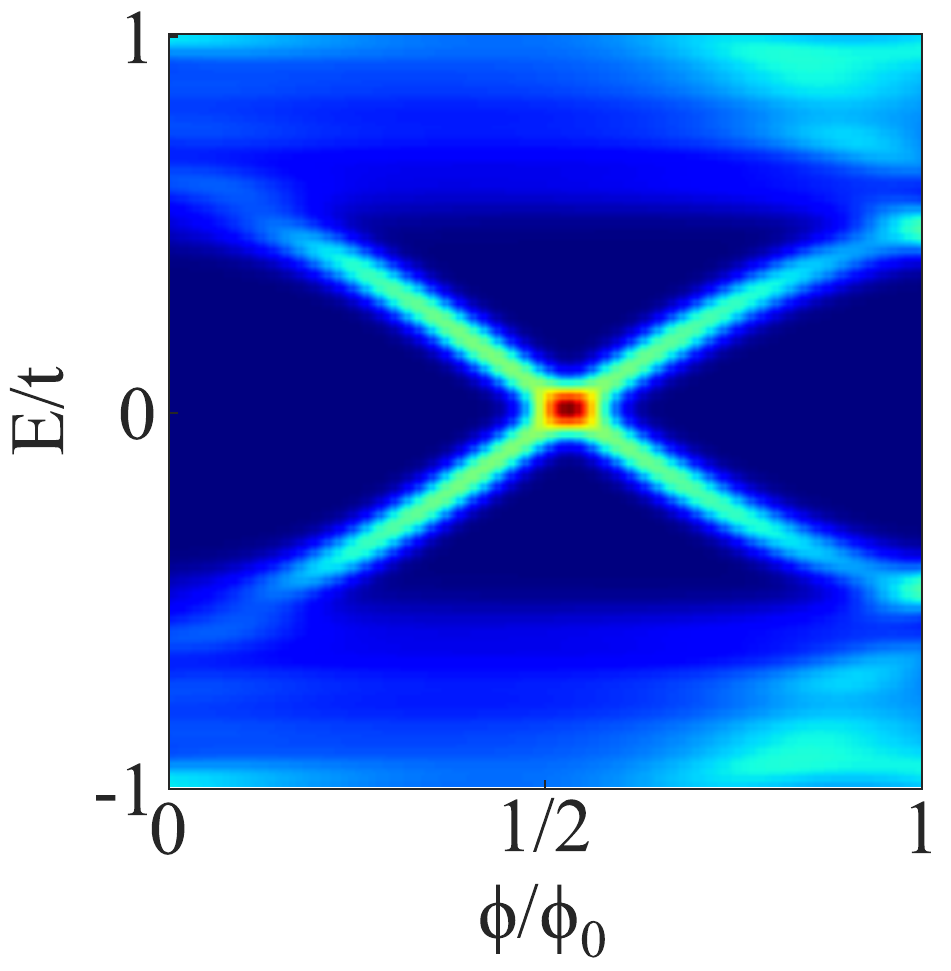}
\label{fig:QuadFlux}}
\subfigure[]{
\includegraphics[scale=0.4]{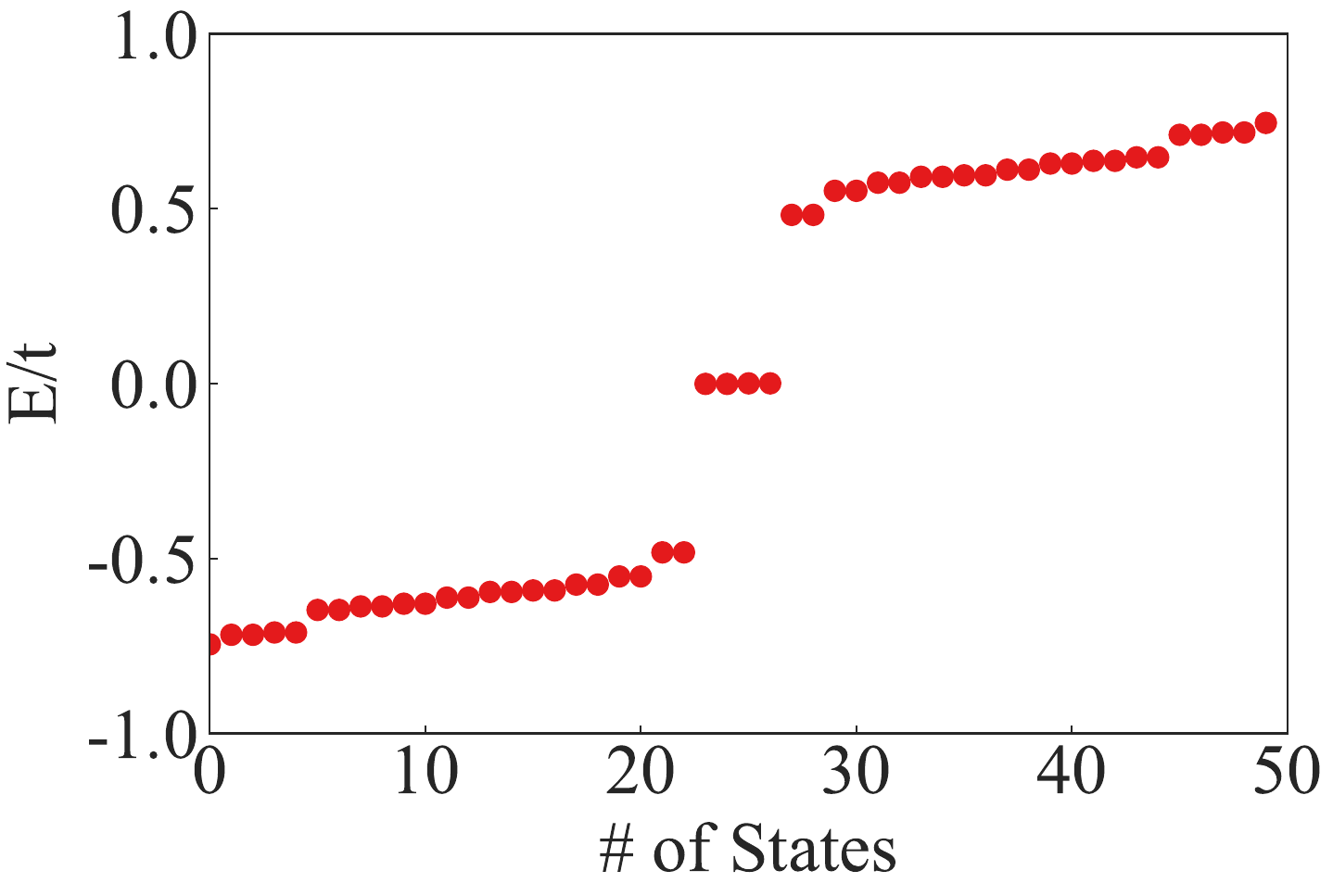}
\label{fig:QuadStates}}
\subfigure[]{
\includegraphics[scale=0.4]{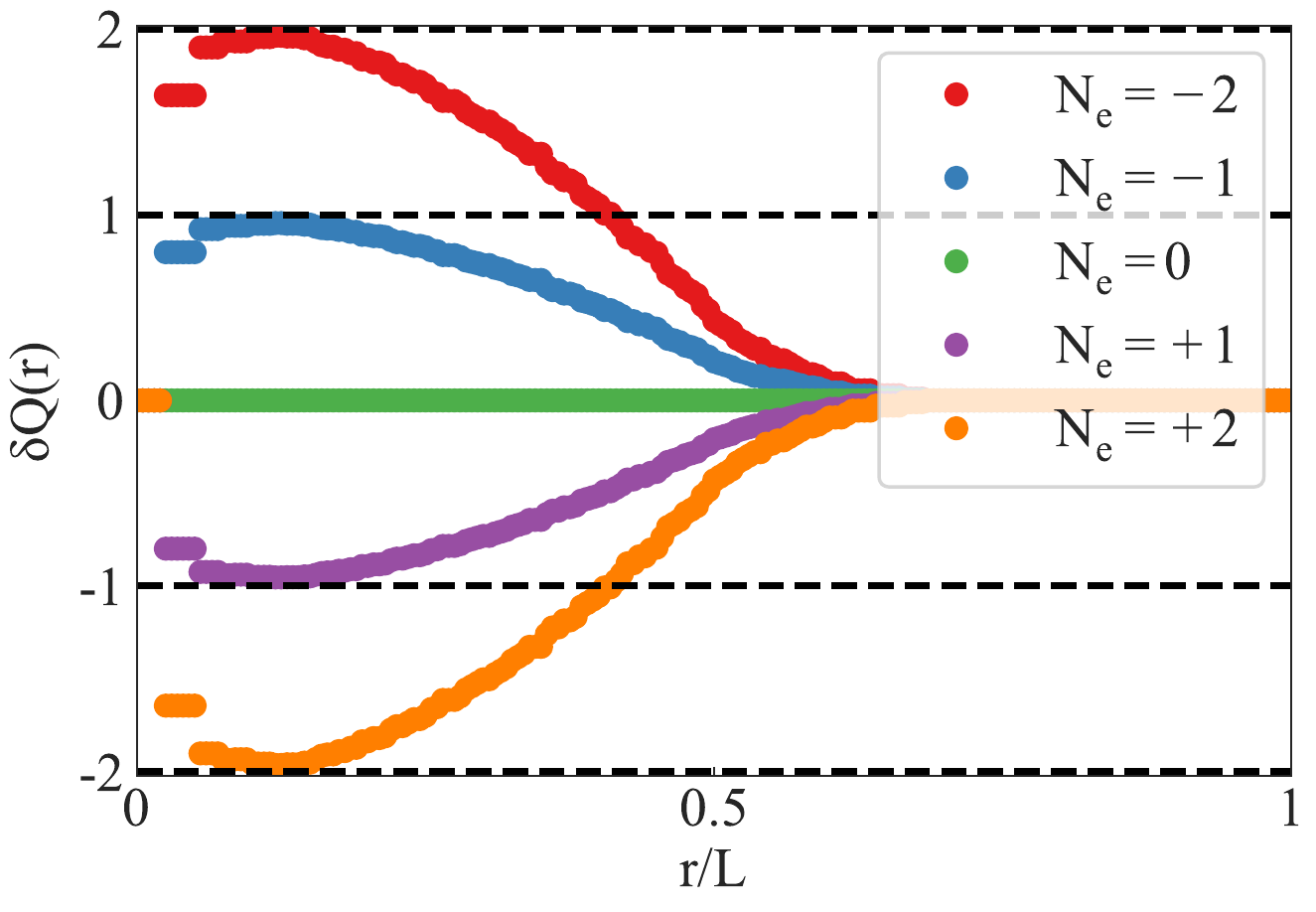}
\label{fig:QuadIC}}
\caption{Spin-charge separation for Phase 2 and Phase 5, possessing $|\mathfrak{C}_{R,GS}|=2 $. All calculations are perfomed for a system size of $24 \times 24$ lattice sites, under periodic boundary conditions.  (a) Local density of states on flux tube as a function of $\phi/\phi_0$. Both branches of spectra, which traverse the bulk gap are two-fold degenerate, implying pumping of two Kramers pairs. (b) At $\phi=\phi_{0}/2$, they lead to four zero-energy bound states, and $6$-fold degeneracy of the half-filled ground state. (c) Induced electric charge (in units of $-e$) on $\pi$-flux tube within a radius $r$, and $N_e$ denotes the number of doped electrons. The maximum induced charge saturates to quantized values $\pm 2e$ for the non-degenerate states ($SU(4)$-singlets), and $ \pm e$ for the four-fold degenerate states ($SU(4)$ quartets), and $0$ for the half-filled state ($SU(4)$ sextet), respectively. }
\label{fig:QuadChern}
\end{figure*}

\section{Spin-charge separation of minimal models}\label{AppA}
Let us consider the following Bernevig-Hughes-Zhang (BHZ) model~\cite{bernevig2006quantum} of $sp$ hybridization on a square lattice
\begin{eqnarray}\label{eq:BHZ}
&&   \frac{ H_{0}(\mathbf{k})}{t}= \sum_{j=1}^{3} d_j(\bs{k} ) \Gamma_j = \sin k_{x}\Gamma_{1}+ \sin k_{y}\Gamma_{2} +  [M  \nn \\ && +\Delta_1(\cos k_{x}  +\cos k_{y}) + \Delta_2 \cos k_x \cos k_y ]\Gamma_{3} ,
\end{eqnarray}
where $\Gamma_{1}=\tau_1\otimes \sigma_{1}$, $\Gamma_{2}=\tau_1\otimes \sigma_{2}$, and $\Gamma_{3}=\tau_3\otimes \sigma_0$ are mutually anti-commuting matrices. The $2\times 2$ identity matrix $\sigma_0$ ($\tau_0$) and Pauli matrices $\sigma_{j=1,2,3}$ ($\tau_{j=1,2,3}$) operate on spin (orbital/parity) index. The hopping parameter $t$ has units of energy, and $M$, $\Delta_1$, $\Delta_2$ are dimensionless tuning parameters, and the lattice constant has been set to unity. The Hamiltonian anti-commutes with $\Gamma_4=\tau_1 \otimes \sigma_3$, and $\Gamma_5=\tau_2 \otimes \sigma_0$, and commutes with $\Gamma_{45}=[\Gamma_4, \Gamma_5]/(2i)=\sigma_3 \otimes \tau_0$. Thus, $\Psi_i^\dagger \mathbb{1} \Psi_i$ and $\Psi_i^\dagger \Gamma_{45} \Psi$ are generators of total number [$U_{+}(1)$] and spin rotation [$U_-(1)$] symmetries, respectively. 

At time-reversal-invariant momentum points $\Gamma: \bs{Q}= (0,0)$, $M: \bs{Q}=(\pi,\pi)$, and $X: \bs{Q}=\{(\pi,0),(0,\pi) \}$ points $H_0 \to t d_3(\bs{Q}) \Gamma_3$, and $[H_0(\bs{Q}), \Gamma_3]=0$. Parity eigenvalues of valence bands are given by $-\text{sgn}(d_3(\bs{Q}))$. A representative phase diagram is shown in Fig.~\ref{fig:phasediagram}, and the pattern of parity eigenvalues and the bulk invariant 
\begin{equation}
\mathfrak{C}_{R,GS}= \frac{1}{4\pi} \int_{BZ} \; d^2k \; \bs{\hat{d}} \cdot \left( \frac{\partial \bs{\hat{d}}}{\partial k_x} \times \frac{\partial \bs{\hat{d}}}{\partial k_y} \right)
\end{equation}
are listed in Table~\ref{tab2}.
As phases 3, 4, 6, and 7 (2 and 5) support $|\mathfrak{C}_{R,GS}|= 1$ ($|\mathfrak{C}_{R,GS}|= 2$), magnetic $\pi$-flux tube would bind $2$ ($4$) zero-energy bound states. Therefore, SCS would be controlled by $SU(2)$ and $SU(4)$ multiplets, respectively (see Fig.~\ref{fig:presentation}). When flux $\phi$ is tuned from $0$ to $\phi_0$, one and two units of spin (Kramers-pair) would be pumped. 

After Fourier transformation, we obtain tight-binding model $H_{0,ij}$ in real-space. In the presence of magnetic flux tube, placed at origin, the matrix elements $H_{0,ij}$ between different lattice sites can be replaced by $H_{0,ij} e^{i \phi_{ij}}$, with $\phi_{ij}= \frac{\phi}{\phi_0} \int_{\bs{r}_i}^{\bs{r}_j} \frac{\hat{z} \times \bs{r}}{\bs{r}^2} \cdot d\bs{l}$. The SCS and spin-pumping for Phases 3, 4, 6, and 7 are controlled by $SU(2)$ multiplets. In Fig.~\ref{4band}, we show the results for Phase 3 and Phase 4. 
The results of SCS governed by $SU(4)$ multiplets are displayed in Fig.~\ref{fig:QuadChern}, clearly showing that $2$ Kramers-pair are being pumped. Due to the enhanced degeneracy of bound states, the maximum induced electric charge can now oscillate between $0,\pm e,\; \text{and}\; \pm 2e$.

There are many ways to break $U_-(1)$ spin rotation symmetry. For example, we can modify $H_j$ as 
 \begin{equation}\label{eq:BHZpert}
H_{0}(\mathbf{k}) \to H(\mathbf{k}) = H_{0}(\mathbf{k})+d_4(\bs{k}) \Gamma_4 + d_5(\bs{k}) \Gamma_5,
 \end{equation}
such that the 2-fold Kramers-degeneracy is preserved. The momentum dependent function $d_4(\mathbf{k})=t_{d,1}(\cos 2k_{x} - \cos 2k_{y})$ and $d_5(\bs{k})= t_{d,2} \sin k_x \sin k_y$ maintain $4$-fold rotation symmetry and also vanish at the time-reversal-invariant momentum points. Following Ref.~\onlinecite{tyner2020topology}, it can be shown that $\mathfrak{C}_{R,GS}$ and SCS for all phases remain unchanged. But $d$-wave perturbations destroy \emph{gapless edge-states}. If $B_{1g}$ term is changed to $(\cos k_x - \cos k_y)$, $X$ points cannot participate in band inversion. Thus, states with even integer winding number become trivialized and no longer display SCS and spin-pumping.
 \begin{figure*}[t]
\centering
\subfigure[]{
\includegraphics[scale=0.7]{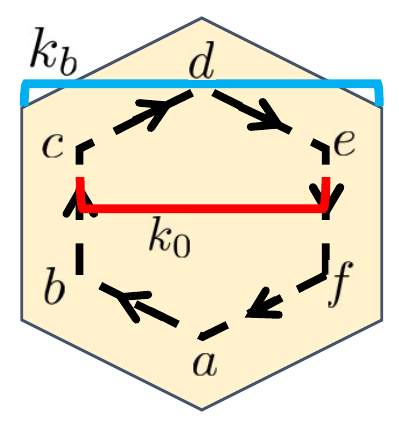}
\label{fig:BiPWL}}
\subfigure[]{
\includegraphics[scale=0.4]{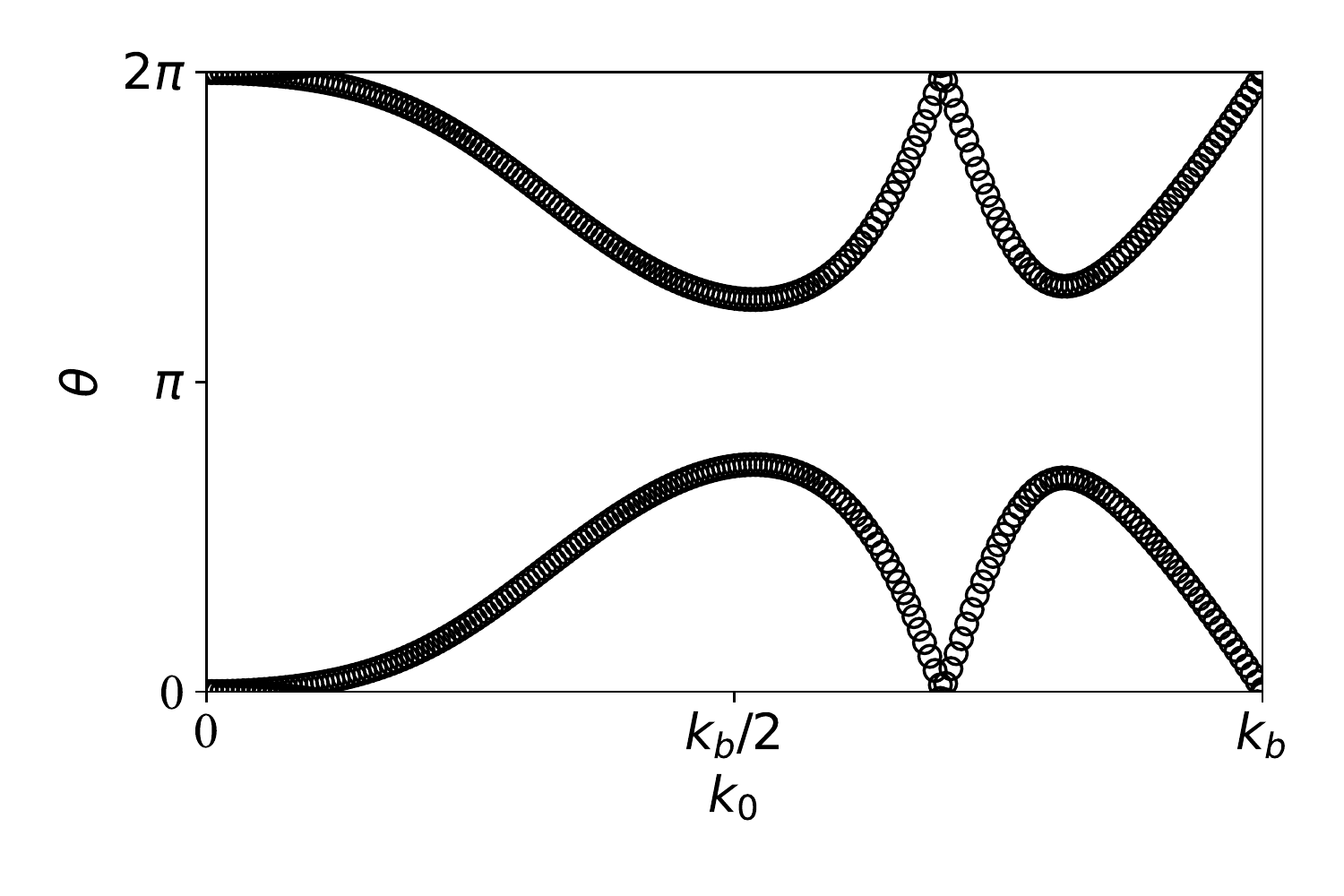}
\label{fig:BiBilayerPWL1}}
\subfigure[]{
\includegraphics[scale=0.4]{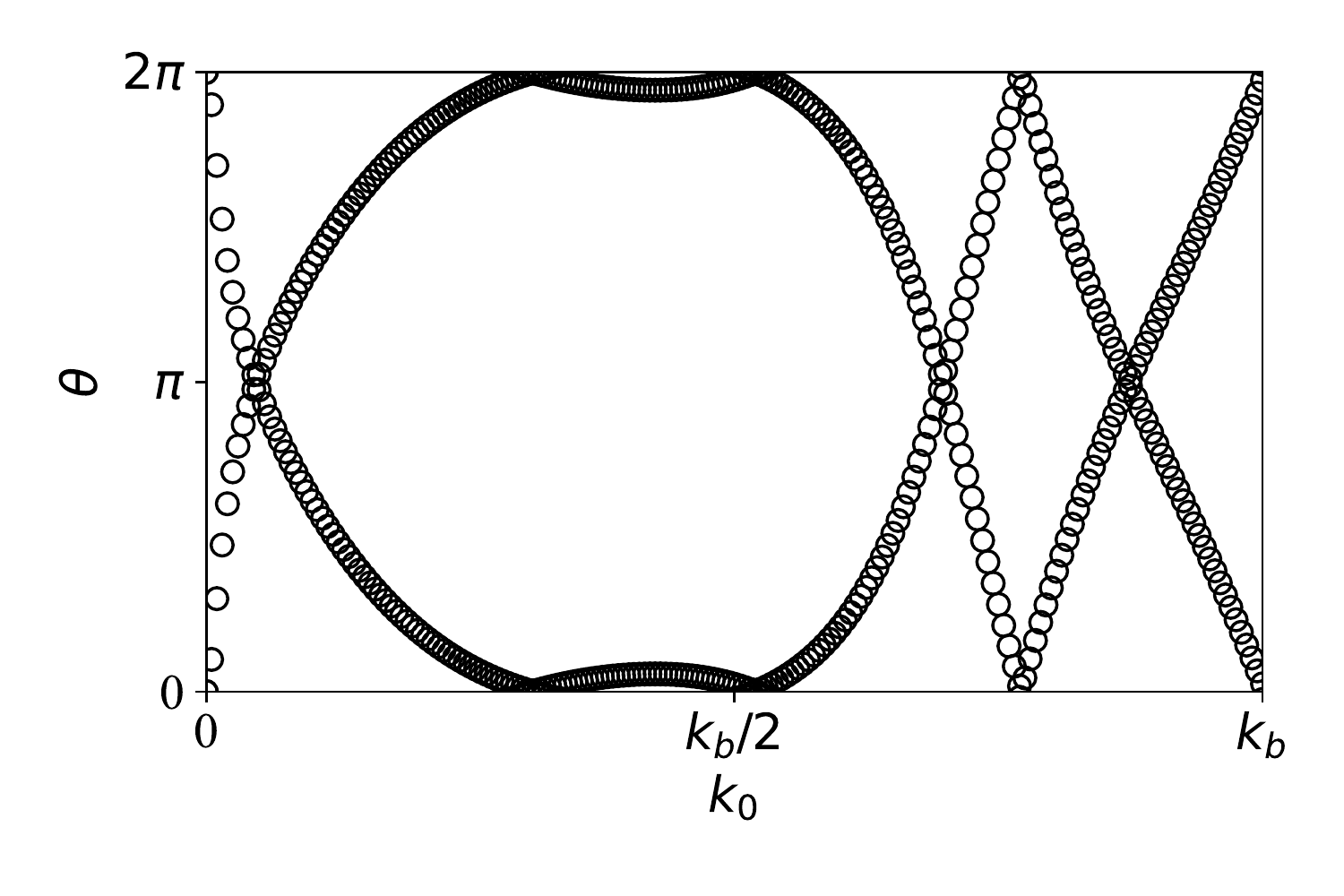}
\label{fig:BiBilayerPWL2}}
\subfigure[]{
\includegraphics[scale=0.4]{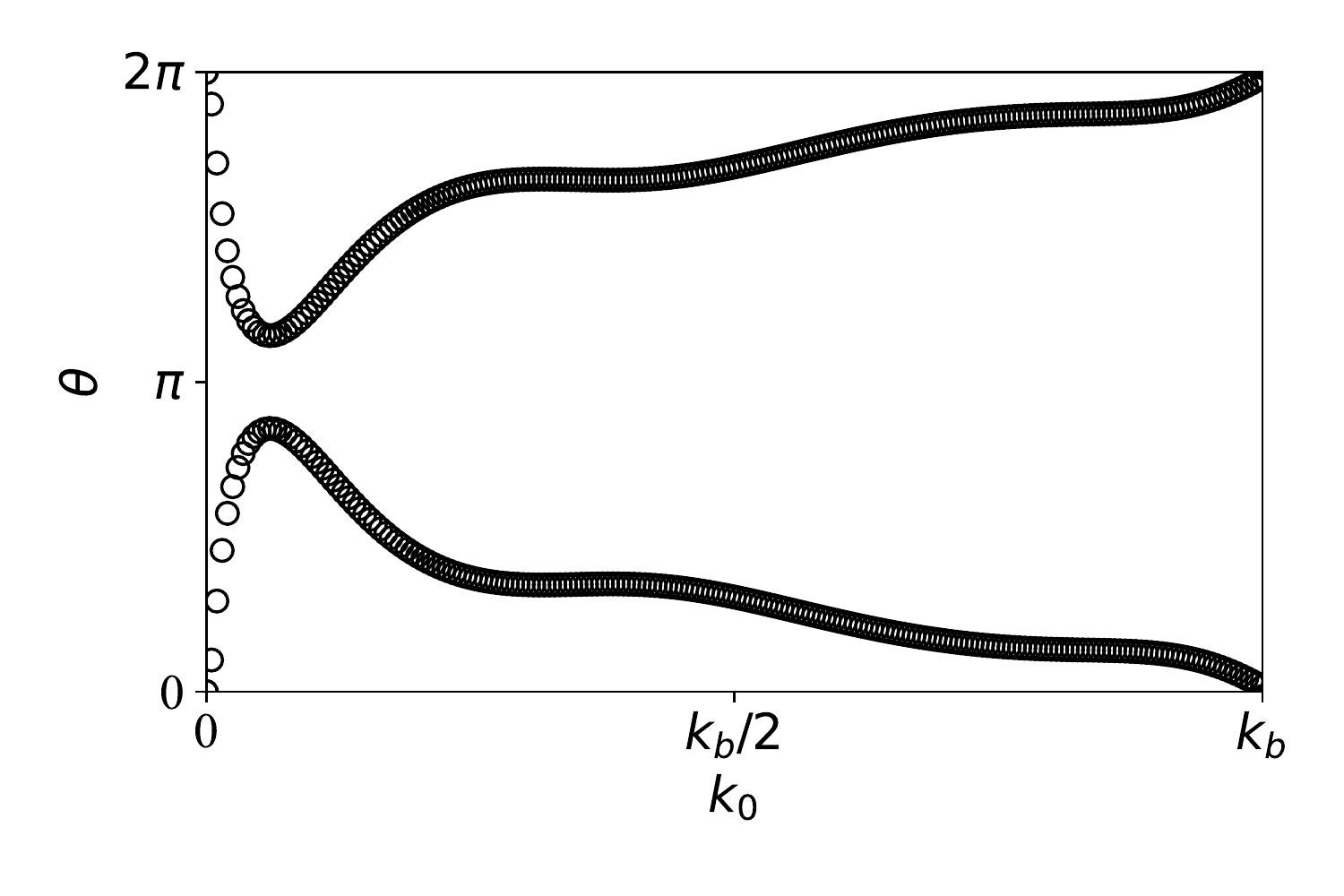}
\label{fig:BiBilayerPWL3}}
\subfigure[]{
\includegraphics[scale=0.4]{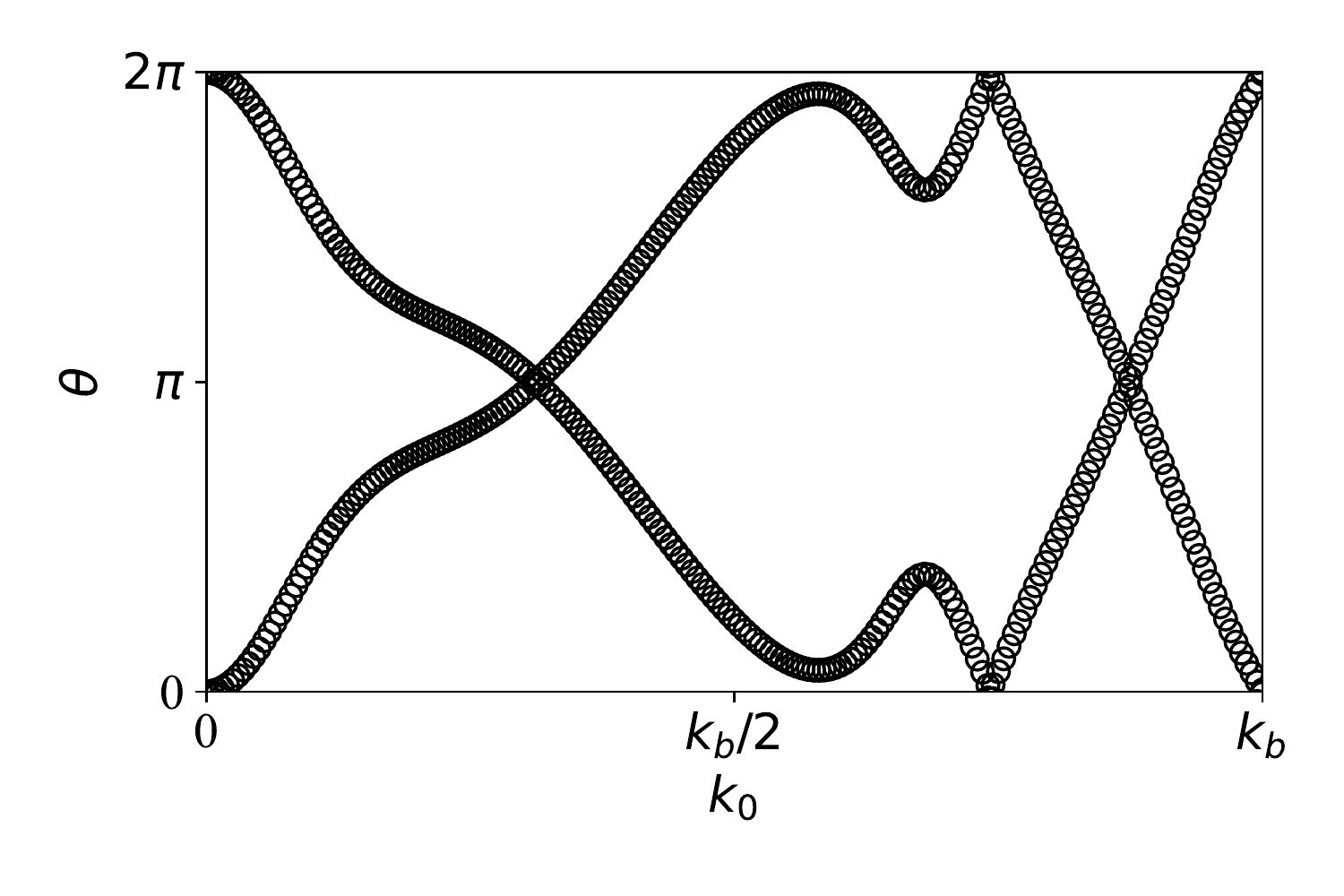}
\label{fig:BiBilayerPWL4}}
\subfigure[]{
\includegraphics[scale=0.4]{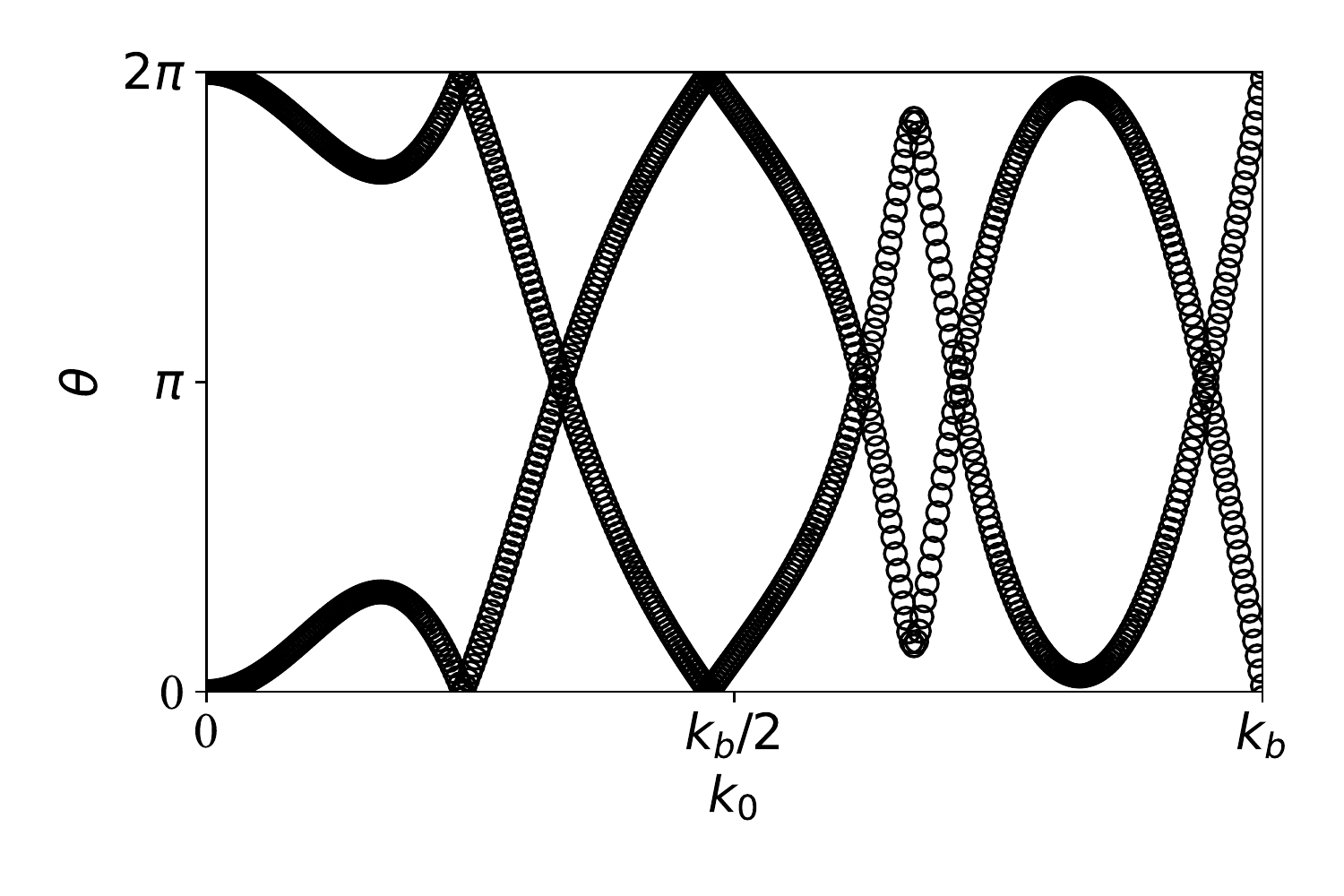}
\label{fig:BiBilayerPWL5}}
\subfigure[]{
\includegraphics[scale=0.4]{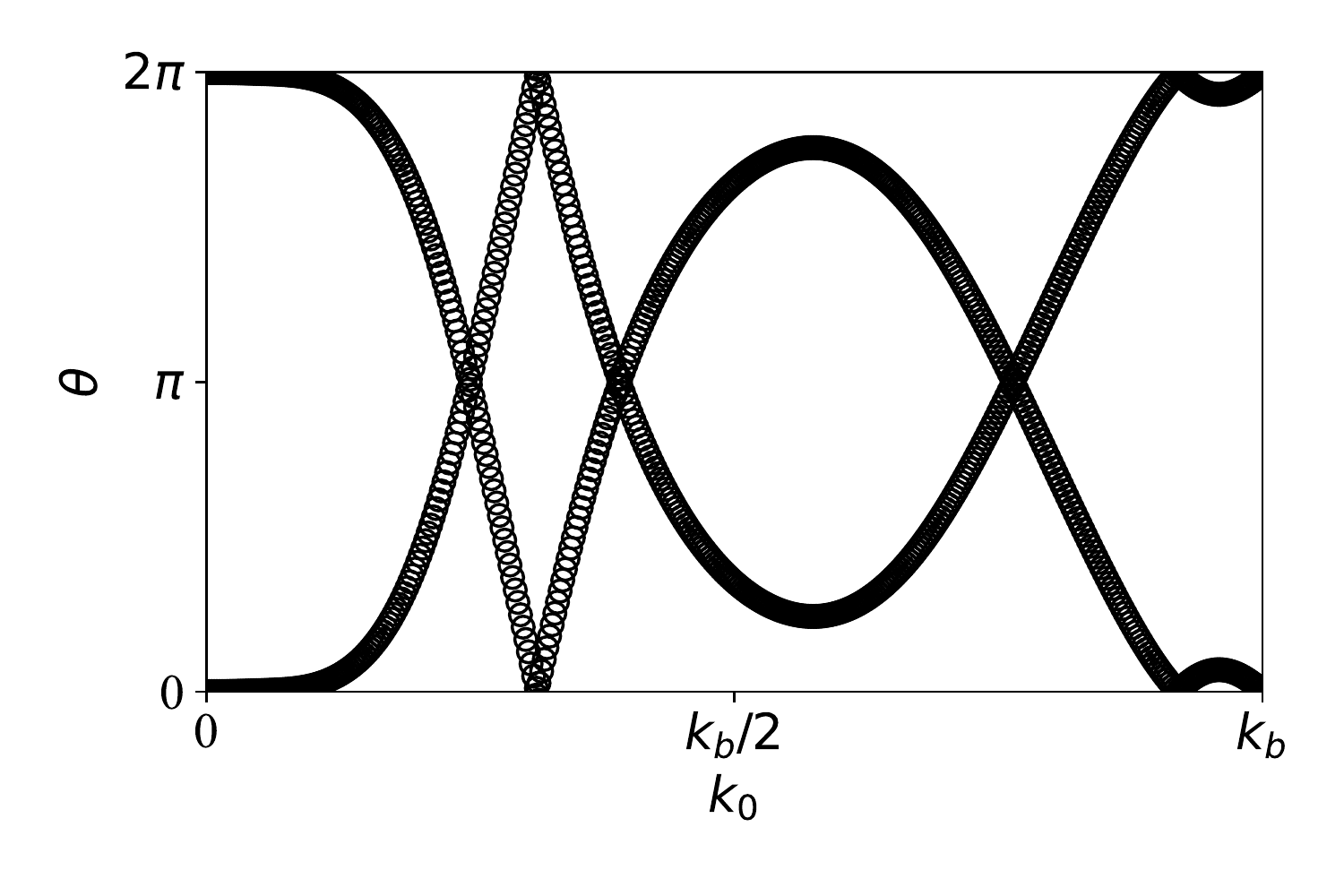}
\label{fig:BiBilayerPWL6}}
\caption{(a) Schematic of path ($abcdef$) for calculating in-plane Wilson loop. The size $k_0$ is increased from $0$ to $k_b$. The results for bands 1-6 are shown in (b)-(g), respectively. Trivial bands 1 and 3 do not show winding of $\theta$. For bands 2 and 6, possessing non-trivial $\mathbb{Z}_2$ index, $\theta$ winds once. For $\mathbb{Z}_2$-trivial bands 4 and 5, $\theta$ winds twice. Therefore, bands 1 through 6 support relative Chern numbers $|\mathfrak{C}_{R,n}|=0,1,0,2,2,1$, respectively, as listed in Table~\ref{tab:Indicators}.}
\end{figure*}

\begin{figure*}[t]
    \centering
  \subfigure[]{  
  \includegraphics[scale=0.65]{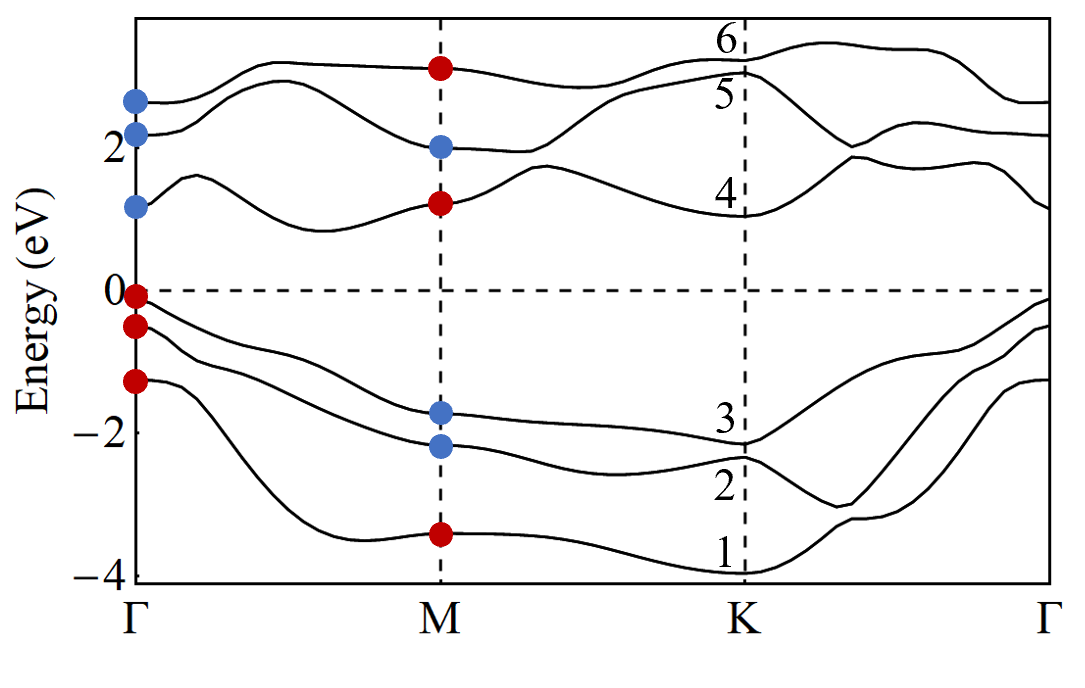}
  \label{Sb1}}
  \subfigure[]{ 
   \includegraphics[scale=0.25]{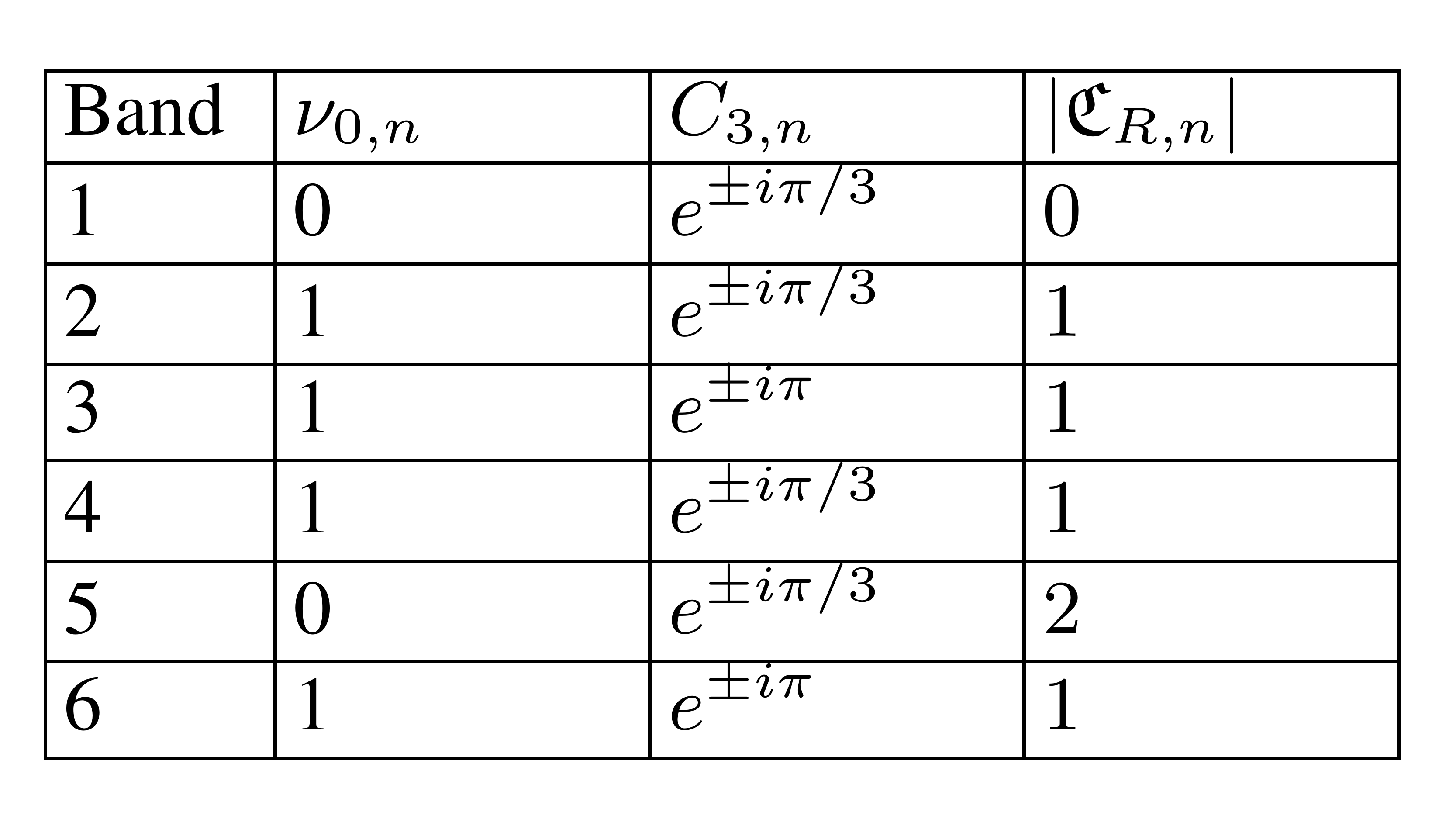}
   \label{Sb2}}
    \subfigure[]{ 
   \includegraphics[scale=0.55]{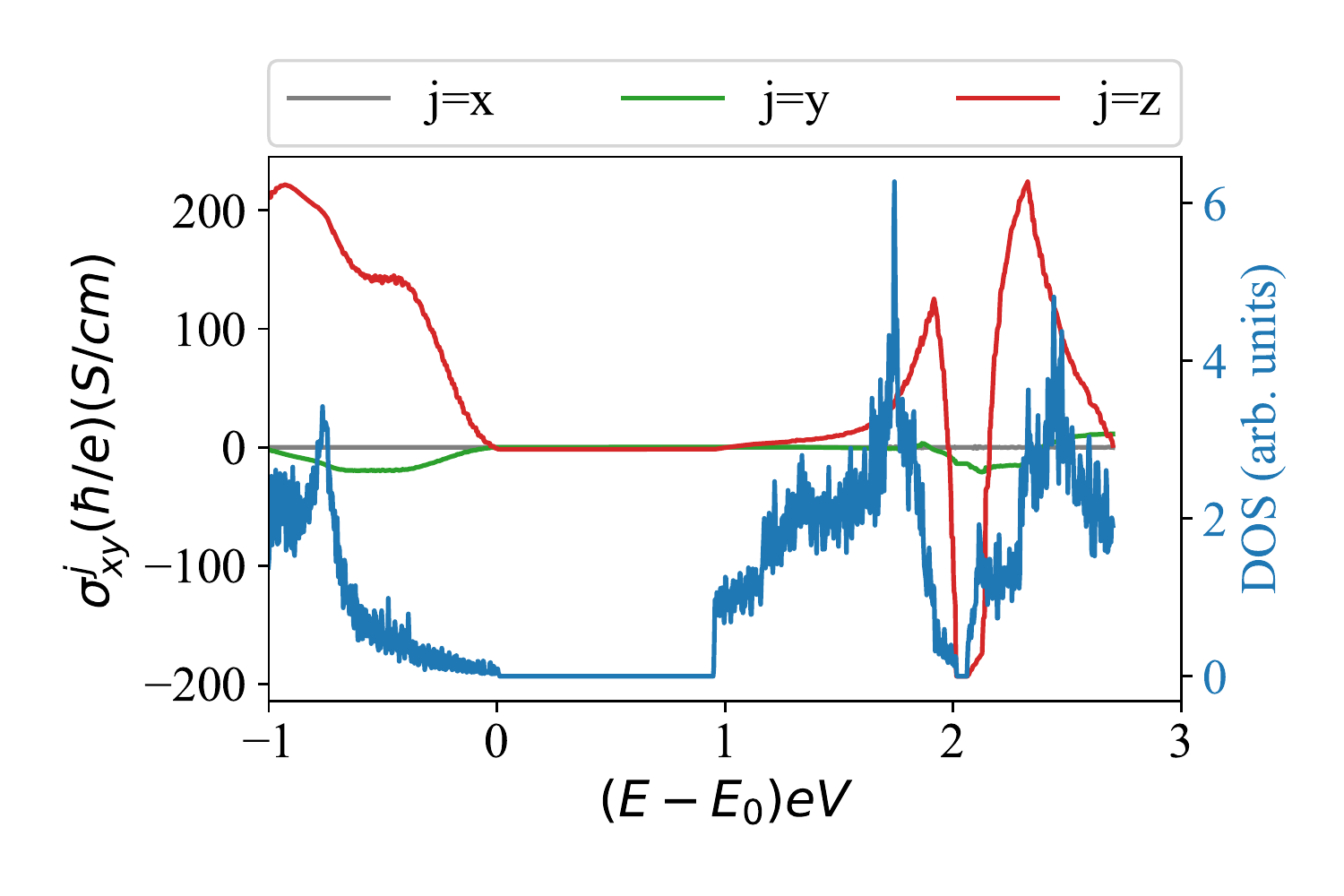}
   \label{Sb3}}
   \subfigure[]{ 
   \includegraphics[scale=0.7]{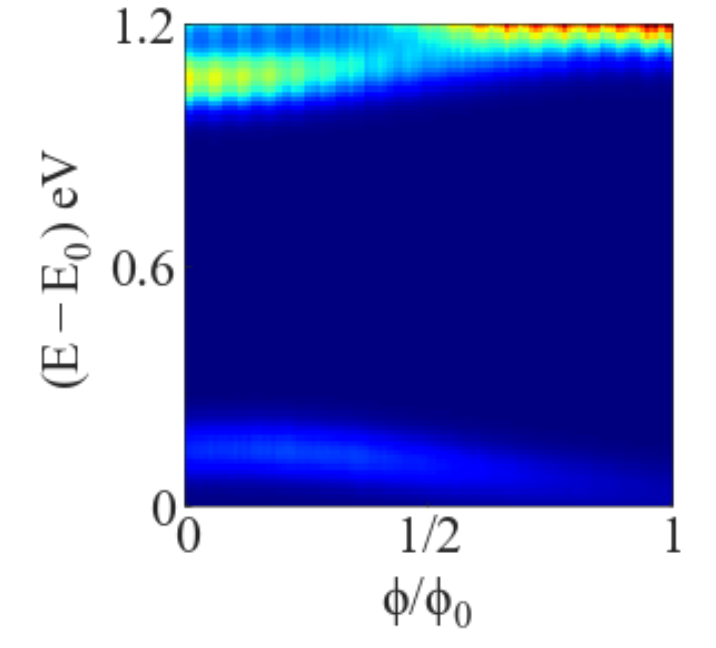}
   \label{Sb4}}
   \caption{(a) Band structure of $\beta$-antimonene, along high-symmetry path of hexagonal Brillouin zone. The bands are numbered according to their energies at $\Gamma$ point, and parity eigenvalue $+1$ ($-1$) at time-reversal-invariant momentum points are denoted by red (blue) dots. (b) Summary of momentum-space topology of constituent bands, where $\nu_{0,n}$, $C_{3,n}$, and $\mathfrak{C}_{R,n}$ respectively denote the $\mathbb{Z}_2$ index, $3$-fold rotation eigenvalue, and the relative Chern number of $n$-th Kramers-degenerate bands. (c) First principle calculations of spin Hall conductivity. When the Fermi level is tuned inside direct band gap, all three components of spin Hall conductivity vanish for the insulating state. (c) Local density of states on the magnetic flux tube in the vicinity bulk band gap does not show any spin-pumping, which shows that the net relative Chern number $\mathfrak{C}_{R,GS}=0$.}
    \label{fig:Sb_Bands}
\end{figure*}

An example of decoupled models with higher number of bands can be found in Ref.~\onlinecite{Wang_2010}. Using a model of three Kramers degenerate bands on the Kagome lattice, with $U(1)$ spin-rotation symmetry, Wang \emph{et. al.} found $(\mathfrak{C}_{R,1}, \mathfrak{C}_{R,2}, \mathfrak{C}_{R,3})=(-1, +2, -1)$. Consequently, $C_{R, GS} = -1, +1$ for $1/3$- and $2/3$- filled insulators, and both states exhibited SCS governed by $SU(2)$ multiplets. This situation is similar to our observations in $\beta$-bismuthene. 

Therefore, we can conclude that flux tube can perform $\mathbb{N}$ and $\mathbb{Z}$ classification of quantum spin Hall states, irrespective of spin-rotation symmetry. We will further substantiate this conclusion by studying momentum space topology and real space response of $(111)$-bilayer of antimony (antimonene) in Sec.~\ref{AppC}.

\section{Magnitude of relative Chern numbers}\label{AppB}
Recently, the in-plane Wilson loop has been utilized to quantify magnitude of $SU(2)$ Berry flux of constituent Kramers-degenerate bands of two-dimensional first and higher-order topological insulators.~\cite{tyner2020topology,tyner2021quantized} The in-plane Wilson loop of $n$-th band measures $SU(2)$ Berry phase accrued upon parallel transport along a non-intersecting closed contour $C$. It is defined by
\begin{equation}
    W_{n}=P \; \text{exp}\left[i\oint A_{j,n}(\mathbf{k})dk_{j}\right]
    =\text{exp}\left(i\theta_{n}\hat{\bs{\Omega}}_n \cdot \mathbf{\sigma}\right),
\end{equation}
where $A^{s s^\prime}_{j,n}(\mathbf{k})=-i \langle \psi_{n,s}(\bs{k}) |\partial_j  \psi_{n,s^\prime} (\bs{k})\rangle$ describes components of $SU(2)$ Berry connection, $\partial_j = \frac{\partial}{\partial k_j}$, $\psi_{n,s=\pm 1}(\mathbf{k})$ are degenerate eigenfunctions of $n$-th band, and $\mathcal{P}$ indicates path-ordering. While the angle $\theta_{n}$ measures gauge-invariant magnitude of non-Abelian flux enclosed by $C$, the three-component unit vector $\hat{\bs{\Omega}}_n$ depends on gauge choice. 

Following the convention of defining gauge-invariant eigenvalues of Wilson lines or Wannier center charges, we analyze eigenvalues of $\text{Im(Ln(}W_{n})) \equiv \pm |\theta_n| \; \text{mod} \; \pi$. In-plane loops are calculated with Wannier90~\cite{Pizzi2020} and Z2Pack software packages, by following $C_3$-symmetry preserving contour, shown in Fig.~\ref{fig:BiPWL}. The area enclosed by the contour is systematically increased from zero to the area of first Brillouin zone. The number of winding of $\theta_n$ corresponds to the absolute value of relative Chern number $|\mathfrak{C}_{R,n}|$. The results for bands 1-6 are shown in Fig.~\ref{fig:BiBilayerPWL1}-\ref{fig:BiBilayerPWL6}.

\section{Analysis of $\beta$-antimonene}\label{AppC}

In contrast to $\beta$-bismuthene, the occupied subspace of single layer of $(111)$ antimony ($\beta$-antimonene) supports trivial $\mathbb{Z}_2$-classification with $\nu_{0, GS}=0$. Whether the ground state supports quantum spin Hall effect can be directly addressed by combined analysis of momentum space topology and real-space response. The bulk band structure and $\mathbb{N}$-classification of constituent bands are shown in Fig.~\ref{Sb1} and Fig.~\ref{Sb2}, respectively. We have used the lattice parameters given by Mounet \emph{et. al.}~\cite{mounet2018two}. Since occupied bands $n=1,2, 3$ possess $|\mathfrak{C}_{R,n}|=0,1,1$, $|\mathfrak{C}_{R,GS}|=0, 2$ are two possible options for the net relative Chern number. 

The first principles calculations show that all three components of spin Hall conductivity vanish for the half-filled insulating state (see Fig.~\ref{Sb3}). To unambiguously probe topological response, we have performed thought experiments with flux tube for a system size of $24 \times 24$ unit cells, under periodic boundary conditions. The spectrum does not show any mid-gap bound states for $N_e=N/2$ and no spin-pumping is observed (see Fig.~\ref{Sb4} ), implying $\mathfrak{C}_{R,GS}=0$. Therefore, $(\mathfrak{C}_{R,2}, \mathfrak{C}_{R,3})=\pm (1, - 1)$ are the possible assignments of signed relative Chern numbers. Due to the lack of any further direct band gaps, we do not pursue the analysis for other filling fractions.

\acknowledgements{
This work was supported by the National Science Foundation MRSEC program (DMR-1720139) at the Materials Research Center of Northwestern University, and the start up funds of P. G. provided by the Northwestern University. A part of this work was performed at the Aspen Center for Physics, which is supported by National Science Foundation grant PHY-1607611.}
\clearpage

\newpage

\bibliographystyle{apsrev4-1}
\nocite{apsrev41Control}
\bibliography{ref.bib}

\end{document}